\documentclass[reprint,
superscriptaddress,
amsmath,
amssymb,
aps,
prb,
floatfix,
english
]{revtex4-2}

\usepackage{bibunits}
\usepackage[unicode=true, colorlinks=true, citecolor={blue!80!black}, urlcolor={blue!50!black}, linkcolor = {blue!80!black}]{hyperref}
\defaultbibliography{references}
\defaultbibliographystyle{apsrev4-2}
\usepackage{graphicx}% Include figure files
\usepackage{svg}
\usepackage{dcolumn}% Align table columns on decimal point
\usepackage{siunitx}
\usepackage[T1]{fontenc}
\usepackage[utf8]{inputenc}
\usepackage{bbm}
\usepackage[normalem]{ulem}

\begin{document}
\widetext

\title{Nonlocal conductance of a Majorana wire near the topological transition}

\author{Vladislav D.~Kurilovich}
\thanks{
Present address: Google Research, Mountain View, CA, USA.\\
Email: vladislav.kurilovich@gmail.com}
\affiliation{Department of Physics, Yale University, New Haven, CT 06520, USA}
\author{William S.~Cole}
\affiliation{Microsoft Quantum, Station Q, Santa Barbara, CA 93111, USA}
\author{Roman M.~Lutchyn}
\affiliation{Microsoft Quantum, Station Q, Santa Barbara, CA 93111, USA}
\author{Leonid I.~Glazman}
\affiliation{Department of Physics, Yale University, New Haven, CT 06520, USA}

\begin{abstract}
We develop a theory of the nonlocal conductance $G_{RL}(V)$ for a disordered Majorana wire tuned near the topological transition critical point. 
{  Under these conditions, the antisymmetric part of the differential conductance, $[G_{RL}(V) - G_{RL}(-V)] /2$, is the dominant one for a sufficiently long wire.
This reflects the charge-neutral nature of the critical modes in the wire.
We factorize the conductance into a term describing propagation of the critical modes along the wire, and terms describing the contacts between the wire and the normal leads.
Topological transition affects only the former term.}
At the critical point, the localization length has a logarithmic singularity at the Fermi level, $l(E) \propto \ln(1 / E)$. This singularity directly manifests in the conductance magnitude, as $\ln |G_{RL}(V) / G_Q| \sim L / l(eV)$ for the wire of length $L \gg l(eV)$. Tuning the wire away from the immediate vicinity of the critical point changes the monotonicity of $l(E)$.
This change in monotonicity allows us to define the width of the critical region around the transition point.

\end{abstract}

\maketitle

\section{Introduction}
A semiconducting nanowire proximitized by a conventional superconductor has recently emerged as a building block for a topological qubit \cite{kitaev2001, lutchyn2010, oreg2010, Karzig2017, MPRpaper2024}.
Magnetic field $B$ applied to such a wire can tune it through a quantum critical point, $B = B_{\rm c}$, into the topological phase.
The defining feature of this phase is the presence of Majorana zero modes (MZMs) localized at the ends of the wire.
MZMs possess an exotic non-Abelian exchange statistics \cite{read2000, ivanov2001}, which gives a pathway to fault-tolerant
%way.
quantum computing \cite{nayak2008, Alicea2012, Karzig2017}. 
The applied and fundamental interest in MZMs
brings about the need for
a robust protocol
allowing one to detect the transition into the topological phase
%has occurred 
in a ``Majorana wire''~\cite{aghaee2023}.

\begin{figure*}[t]
  \begin{center}
    \includegraphics[scale = 1]{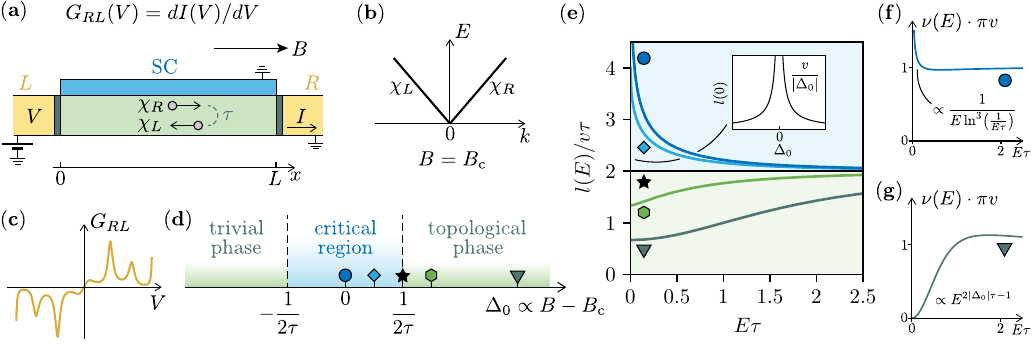}
    \caption{
    {\bf a.} 
    Sketch of the considered setup. A semiconducting wire (green) is proximitized by a superconductor (blue) and contacted at its ends by normal leads (yellow). Magnetic field $B$ tunes the wire to the vicinity of the topological phase transition.
    We find the nonlocal differential conductance $G_{RL}(V) = dI(V) / dV$; $V$ is the bias applied to the left lead and $I$ is the current collected at the right lead. {\bf b.} At the critical point, $B = B_{\rm c}$, the low-energy degrees of freedom in the wire are a pair of counter-propagating Majorana fermions with linear dispersion relation $E(k)$. {\bf c.} Neutral character of Majorana fermions renders $G_{RL}(V)$ an odd function at low biases. {\bf d.} Phase diagram of the Majorana wire. $\Delta_0 \propto B - B_{\rm c}$ characterizes the detuning from the critical point.  Positive and negative values of $\Delta_0$ correspond to the topological and trivial phase, respectively.
    The critical region is the range of fields for which $|\Delta_0| \leq 1 / 2 \tau$; its width is determined by the Drude mean free time $\tau$.
    The properties of the wire differ qualitatively inside and outside of the critical region, as illustrated in panels {\bf e}--{\bf g}.
    {\bf e.} Evolution of the energy dependence of the localization length $l(E)$ with $B$. At the critical point, $l(E)$ diverges logarithmically at $E \rightarrow 0$ [see Eq.~\eqref{eq:log_div_intro}]. The monotonicity of $l(E)$ changes as the wire is tuned across the boundary of the critical region. Inset: field-dependence of the localization length at the Fermi level [see Eq.~\eqref{eq:loc_at_FL_intro}]. {\bf f.} In the critical region, the density of states is singular at $E = 0$. {\bf g.}~Outside of the critical region, the singularity is replaced by a ``soft'' gap, $\nu (0) = 0$.}
    \label{fig:summary}
  \end{center}
\end{figure*}

An early approach to the topological phase identification relied on the local transport measurements.
It was guided by the prediction that MZMs lead to zero-bias peaks in the \textit{local} conductances measured at the wire's ends \cite{sengupta2001,Bolech2007, Nilsson2008, law2009,Flensberg2010, sau2010,Stanescu2011,Fidkowski2012}.
The zero-bias peaks were indeed observed experimentally \cite{mourik2012,Lutchyn2018}. However, Majorana zero-bias peaks can be mimicked by resonances of a trivial nature, such as those produced by disorder \cite{lee2012, haining2020, dassarma2021}, the Kondo physics \cite{ejhlee2012, Meng2014}, geometrical resonances leading to “quasi-Majorana” modes~\cite{vuik2019}, or other types of trivial Andreev levels~\cite{reeg2018}.
To reduce the probability of false positives, Refs.~\cite{puglia2021, pikulin2021, aghaee2023, banerjee2023} proposed complementing local measurements of zero-bias peaks with a measurement of nonlocal conductance, see Fig.~\ref{fig:summary}(a). 
The latter quantity,  $G_{RL}(V)$, is largely determined by the bulk properties of the wire, and, therefore, is less susceptible to the imperfections at the wire's ends. 
Ideally, the nonlocal conductance measurements should detect the gap closing and re-opening 
%required 
at the topological phase transition~\cite{read2000, rect2018, Danon2020,Maiani2022}. %, see Fig.~\ref{fig:summary}(a). 
The combination of stable zero-bias peaks and gap closing and re-opening in the region of interest significantly reduces the probability of false positive scenarios~\cite{aghaee2023}, notwithstanding the discussion in Ref.~\onlinecite{haining2021, hess2023, dassarma2023}.

Indeed, in a clean wire, the phases on the two sides of the critical point both have a spectral gap $E_{\rm gap}$, while at the critical point itself the gap closes. 
A quasiparticle with $E > E_{\rm gap}$ can propagate along the wire freely, whereas a quasiparticle with $E < E_{\rm gap}$ can only propagate as an evanescent wave. Thus, $G_{RL}(V)$ of a long wire is appreciable only above the threshold voltage, $V > E_{\rm gap}/ e$. This gives a direct electrical access to $E_{\rm gap}$. The closing and re-opening of the gap allows one to pinpoint the critical field value $B_{\rm c}$ of a clean wire.

In contrast to the ideal scenario, disorder in real devices makes the connection between $G_{RL}(V)$ and the properties of the topological phase more complex.
However, apart from numerical modeling \cite{Stanescu2011, rect2018, pikulin2021,  haining2021, aghaee2023, dassarma2023}, this connection has not been studied.
In this work, we develop an analytical theory for the nonlocal conductance of a disordered, finite-length wire near the critical point.

The presence of finite disorder in the wire does not destroy the topological phase, provided the superconducting coherence length in the topological phase is shorter than the normal-state localization length~\cite{motrunich2001, gruzberg2005, brouwer2011, Lobos2012, DeGottardi_thesis, adagideli2014effects}. However, disorder modifies the character of the phase transition~\cite{fisher1994}. Because the time-reversal symmetry is broken by the field, the Anderson theorem does not apply. Disorder leads to the formation of subgap states, thus closing the spectral gap  
of “dirty” Majorana wires \cite{brouwer2011, brouwer2011-2}. In fact, the modification of the wire’s density of states (DOS) by disorder is most prominent near the critical point \cite{motrunich2001, brouwer2011-2}. There, the DOS is singular, diverging at the Fermi level, $E \rightarrow 0$. The modification of DOS smears the sharp threshold behavior of $G_{RL}(V)$ characteristic of the clean limit.

To elucidate the universal features in $G_{RL}(V)$ associated with the topological phase transition, we 
separate the effects of the quasiparticle propagation along a disordered wire from the effects of the contacts.
The ability of quasiparticles to propagate is encoded in their energy-dependent localization length $l(E)$.
We develop a theory for the behavior of $l(E)$ in the vicinity of the topological transition, and show how $l(E)$ can be extracted from the measurement of $G_{RL}(V)$.
We find good agreement between the universal low-energy theory and numerical simulations for the Rashba wire model. 

\subsection*{Summary of the results}
We develop a theory of nonlocal transport across a disordered Majorana wire of {length $L$, tuned to the vicinity of} the topological transition,  $|B - B_{\rm c}| \ll B_{\rm c}$.
There, the physics of the wire allows for a universal description.
The low-energy degrees of freedom at $B = B_{\rm c}$ are a pair of counter-propagating Majorana modes, see Figs.~\ref{fig:summary}(a, b). 
The modes are hybridized by detuning of the field from its critical value as well as by disorder.
The neutral character of the Majorana modes is directly reflected in the bias dependence of the conductance. 
It renders $G_{RL}(V)$ an \textit{odd} function at low biases, see Fig.~\ref{fig:summary}(c).
Specifically, we find that $G_{RL}(V)$ factorizes in a product of three terms:
\begin{align}\label{eq:G_RL_P_intro}
    G_{RL}(V) &/ G_Q \notag \\
    &= \bigl(T^R_{{\rm e m}}(E) - T^R_{{\rm h m}}(E)\bigr) {T_{{\rm me}}^L(E) Q(E)}\bigr|_{E = eV}.
\end{align}
Here $T_{\rm me}^L(E)$ describes the conversion of an electron impinging on the wire from the left lead into a Majorana fermion. The Majorana fermion is converted into an electron or a hole at the junction with the right lead. The respective conversion probabilities are $T_{\rm em}^R(E)$ and $T_{\rm hm}^R(E)$. Application of the particle-hole transformation demands that the product of the first two terms in Eq.~\eqref{eq:G_RL_P_intro} is odd-in-$E$ close to the Fermi level. Finally, factor $Q(E)$ describes propagation of a Majorana fermion from one end of the wire to another end. This is an even-in-$E$ function which is determined by disorder in the wire, detuning from the critical point $B - B_{\rm c}$, and the wire length $L$.

{Both disorder} and detuning from the critical point lead to localization of wavefunctions in the wire; this makes $G_{RL}(V)$ 
exponentially small in $L$. A useful quantity to characterize the nonlocal transport is thus the logarithm of conductance. 
In contrast to $G_{RL}(V)$, $\ln |G_{RL}(V) / G_Q|$ is self-averaging, i.e., its value in a given device is representative of an ensemble of devices with different disorder realizations [we normalized $G_{RL}(V)$ by the conductance quantum $G_Q = e^2 / h$]. 
The conductance logarithm satisfies
\begin{equation}
\langle \ln \bigl|G_{RL}(V) / G_Q| \rangle = - \frac{2L}{l(E)}\Bigr|_{E = eV},
\end{equation}
where  $\langle \dots \rangle$ denotes the ensemble average and 
$l(E)$ is the localization length [we disregard an $L$-independent contribution coming from the first two terms in Eq.~\eqref{eq:G_RL_P_intro}]. 
We find the energy-dependence of $l(E)$ at and in the vicinity of the critical point.

The critical point of a disordered Majorana wire, $B = B_{\rm c}$, belongs to the infinite-randomness universality class \cite{motrunich2001, fisher1994}.
The peculiar character of this criticality is reflected in the singular, logarithmic behavior of the localization length. In Sec.~\ref{sec:disorder_crit} we find:
\begin{equation}\label{eq:log_div_intro}
    l(E) = v\tau \ln \Bigl(\frac{1}{E\tau}\Bigr). 
\end{equation}
We expressed $l(E)$ in terms of the velocity $v$ of the Majorana fermions in the wire and Drude relaxation time $\tau$. The latter has the meaning of the mean free time in the absence of interference. 
The divergence of  $l(E)$ at the Fermi level reflects the appearance of a delocalized mode at the topological transition \cite{akhmerov2011}.

\begin{figure}[t]
  \begin{center}
    \includegraphics[scale = 1]{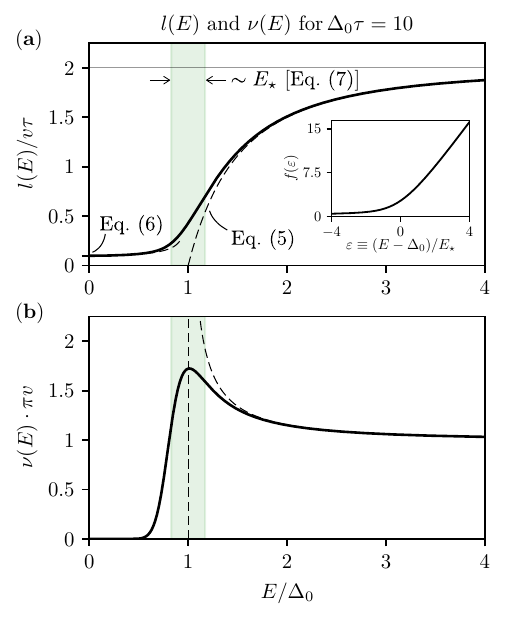}
    \caption{{\bf a.} Energy dependence of the localization length $l(E)$ for a wire tuned away from the critical point by $\Delta_0 \gg 1 / \tau$ (solid line). Dashed lines are the asymptotes for the subgap and above-the-gap behavior of $l(E)$. In the crossover region (shaded green), the dependence of $l(E)$ on $E  - \Delta_0$, $\Delta_0$, and $1/\tau$ has a scaling form [see Eqs.~\eqref{eq:scaling} and \eqref{eq:scaling_2}]. Insets shows the scaling function $f(\varepsilon)$.
    {\bf b.} Density of states $\nu (E)$ plotted with the help of Eq.~\eqref{eq:dos_full} (solid line). Dashed line is the respective result for a clean wire [Eq.~\eqref{eq:dos_0}]. Disorder smears the ``coherence'' peak but a relatively well-pronounced spectral ``gap'' remains for $\Delta_0 \gg 1 / \tau $.}
    \label{fig:loc_away_intro}
  \end{center}
\end{figure}

Tuning the wire away from the critical point removes the divergence of $l(E)$ at the Fermi level. Instead, the localization length saturates at $E \rightarrow 0$ at a constant value
\begin{equation}\label{eq:loc_at_FL_intro}
    l(0) = \frac{v}{|\Delta_0|} \propto \frac{1}{|B - B_{\rm c}|},
\end{equation}
where $\Delta_0 \propto B - B_{\rm c}$ is the energy scale characterizing the deviation from the critical point. 

The energy dependence of $l(E)$ has a different character in the immediate vicinity of the critical point and away from it. We define the boundary of the \textit{critical region} by the condition $|\Delta_0| \tau = 1 / 2$ [see Fig.~\ref{fig:summary}(d)]. Throughout the critical region, $l(E)$ is the largest at the Fermi level, and monotonically decreases away from it, see Fig.~\ref{fig:summary}(e). The trend is opposite outside of the critical region, $|\Delta_0| \tau > 1/2$. There, $l(E)$ is the shortest at $E=0$ and increases monotonically with $|E|$. By investigating the evolution of the $l(E)$ dependence with $B$, one may identify the critical region and locate the topological phase transition. We note that the width of the critical region yields information about the disorder strength.

In addition to the localization length $l(E)$, the criticality is reflected in the low-energy behavior of the wire's density of states (DOS) $\nu(E)$. At $B = B_{\rm c}$, the DOS has a Dyson singularity, $\nu(E) \propto 1/(E \ln^3(1/E\tau))$ \cite{dyson1953, motrunich2001}, see Fig.~\ref{fig:summary}(f). This is a the hallmark feature of the infinite-randomness critical point. 
Tuning $B$ away from $B_{\rm c}$ replaces the Dyson singularity by a milder power-law energy-dependence, $\nu(E) \propto 1 / E^{1 - 2|\Delta_0|\tau}$.
Still, the DOS remains divergent at $E \rightarrow 0$ within the critical region, $|\Delta_0 |\tau < 1/2$. Outside of the critical region, $|\Delta_0| \tau > 1/2$ the divergence is replaced by a ``soft'' gap with $\nu(0) = 0$, see Fig.~\ref{fig:summary}(g).

One can establish the behavior of $l(E)$ outside of the critical region analytically for a large detuning $|\Delta_0| \tau \gg 1$. In this regime, the spectrum of the wire has a relatively well-established gap of magnitude $E_{\rm gap} = |\Delta_0|$, see Fig.~\ref{fig:loc_away_intro}(b). At the above-the-gap energies, $E > |\Delta_0|$, the localization length is given by
\begin{equation}\label{eq:above_intro}
    l(E) = 2v\tau \Bigl(1 - \frac{\Delta_0^2}{E^2}\Bigr).
\end{equation}
It reaches a constant $l(E) \approx 2 v \tau$ at  $E \gg |\Delta_0|$, and decreases linearly as $E$ approaches the gap edge, $l(E) \approx 4v\tau (E - |\Delta_0|) / |\Delta_0|$.
At the subgap energies, $E < |\Delta_0|$, the localization length becomes independent of disorder; we find
\begin{equation}\label{eq:below_intro}
    l(E) = \frac{v}{\sqrt{\Delta_0^2 - E^2}}.
\end{equation}
The crossover between asymptotes \eqref{eq:above_intro} and \eqref{eq:below_intro} happens over a narrow region of width $E_\star \ll |\Delta_0|$ around $E = |\Delta_0|$, see Fig.~\ref{fig:loc_away_intro}(a). In Sec.~\ref{sec:disorder_away}, we obtain
\begin{equation}
    E_\star \sim |\Delta_0|\bigl/(|\Delta_0|\tau\bigr)^{2/3}.
\end{equation}
The crossover is described by  $l(E) = v \tau \cdot (E_\star / |\Delta_0|) \cdot f([E - |\Delta_0|]/E_\star)$, see Fig.~\ref{fig:loc_away_intro}(a) and Eq.~\eqref{eq:scaling_2} for the form of the scaling function $f(\varepsilon)$.

The critical behavior of $l(E)$ is the same on the two sides of the topological transition point. 
Nonetheless, the nonlocal transport measurements still carry an imprint of whether the phase is topological or trivial. 
We find that the Majorana zero modes present in the topological phase lead to a low-bias peak in the nonlocal differential \textit{noise}, see Eq.~\eqref{eq:noise_resonant}. A respective signature in the conductance is strongly suppressed by the neutral character of Majorana fermions, see Eq.~\eqref{eq:resonant} and the related discussion.

The manuscript is organized in the following way.
We introduce the universal low-energy theory of a Majorana wire tuned to a vicinity of the critical point in Sec.~\ref{sec:low-energy_theory}. We use it to find the general expression for the nonlocal conductance [see Eqs.~\eqref{eq:G_RL_P} and \eqref{eq:PE_general}].
As a warm-up, we first apply the theory to describe the transport across a clean wire both at $B = B_{\rm c}$ [see Sec.~\ref{sec:clean_crit}]  and in its vicinity [see Secs.~\ref{sec:clean_away} and \ref{sec:majoranas}].
We then account for disorder in Sec.~\ref{sec:disorder}. We find the conductance of a disordered wire at the critical point in Sec.~\ref{sec:disorder_crit}, and away from it in Secs.~\ref{sec:disorder_away} and \ref{sec:disorder_strong}. We conclude by illustrating how our critical theory applies to a concrete microscopic model of a Majorana wire, see Sec.~\ref{sec:micro}.

\section{Low-energy theory of nonlocal transport\label{sec:low-energy_theory}}
\subsection{The model}
We consider a three-terminal device schematically depicted in Fig.~\ref{fig:summary}(a).
A semiconducting quantum wire is proximitized by a grounded superconductor.
The two ends of the wire are connected to normal metal leads allowing for the  nonlocal transport measurements.
The Hamiltonian of this setup reads:
\begin{equation}\label{eq:H}
    H = H_{\rm wire} + H_L + H_R.
\end{equation}
Here $H_L$ and $H_R$ describe the left and right leads, respectively, as well as their coupling to the wire.
For now, we leave $H_{R / L}$ unspecified. 
The term $H_{\rm wire}$ in Eq.~\eqref{eq:H} is the Hamiltonian of the electronic degrees of freedom in the wire. 

Application of magnetic field $B$ to the wire tunes it across the quantum critical point into the  topological phase. 
We focus on the vicinity of this topological transition.
The physics there admits a universal description {insensitive} to the microscopic details of the system \cite{kurilovich2021}.

To begin with, let us consider a clean wire tuned exactly to the critical point, $B = B_{\rm c}$. In this case, the spectral gap in the wire closes. The only low-energy degrees of freedom are a single pair of counter-propagating Majorana modes with linear energy dispersion. {We consider the limit of a long wire, and therefore dispense with the contribution of higher-energy, non-critical modes to the transport.} The critical modes are described by the following effective Hamiltonian:
\begin{equation}\label{eq:wire_c}
    H_{\rm wire, c} = -\frac{i}{2}\int dx\,\bigl(v_R\chi_R \partial_x \chi_R - v_L\chi_L \partial_x \chi_L\bigr).
\end{equation}
Here $\chi_R(x)$ and $\chi_L(x)$ are Majorana field operators corresponding to right- and left-moving modes, respectively. The anticommutation relation for the field operators is $\{ \chi_\alpha(x), \chi_\beta(x^\prime) \} = \delta_{\alpha\beta} \delta(x - x^\prime)$.
The propagation velocities $v_R$ and $v_L$ may be different in the general case. In the following, though, we will assume that $v_R = v_L \equiv v$.
This situation is realized naturally if the wire has special symmetries such as the inversion symmetry (see Ref.~\onlinecite{kurilovich2021} for further detailed discussion).
The energy dispersion of the Majorana modes then is $E_{R / L}(k) = \pm v k$, where $k$ is the wave vector, see Fig.~\ref{fig:summary}(b).

Tuning $B$ away from its critcal value $B_{\rm c}$ leads to the hybridization of right- and left-moving Majorana modes. In fact, there is only one relevant (in the renormalization group sense) bilinear operator describing such a hybridization:
\begin{equation}\label{eq:detuning}
    \delta H_{\rm wire} = i\Delta_0 \int dx
\, \chi_R(x) \chi_L(x).
\end{equation}
Parameter $\Delta_0$ here is determined by the detuning from the critical point, $\Delta_0 \propto B - B_{\rm c}$. In a clean wire, hybridization described by Eq.~\eqref{eq:detuning} results in the opening of a spectral gap $E_{\rm gap} = |\Delta_0|$. 

Disorder in the wire also leads to the coupling between the two species of Majorana modes. In the low-energy theory, the disorder is described by a position-dependent hybridization term:
\begin{equation}\label{eq:disorder}
    H_{\rm dis} = i \int dx\,\delta \Delta(x) \chi_R(x) \chi_L(x).
\end{equation}
Because the form of the bilinear is the same in Eqs.~\eqref{eq:detuning} and \eqref{eq:disorder}, the average value of the disorder potential $\delta \Delta(x)$ can be absorbed into the shift of the critical field value $B_{\rm c}$. Thus we assume $\langle \delta \Delta(x) \rangle = 0$ without loss of generality. We will specify statistics of $\delta \Delta(x)$ in Sec.~\ref{sec:disorder}.

In total, the Hamilontian of the wire reads
\begin{equation}\label{eq:H_wire}
    H_{\rm wire} = H_{\rm wire, c} + \delta H_{\rm wire} + H_{\rm dis},
\end{equation}
where the three terms are given by Eqs.~\eqref{eq:wire_c}, \eqref{eq:detuning}, and \eqref{eq:disorder}, respectively.
On a single-particle level, this Hamiltonian is identical to the random-mass Dirac model \cite{ovchinnikov1977, brouwer2011}. 
Below we study its transport properties.

\subsection{General expression for the nonlocal conductance\label{sec:general}}
We are interested in the nonlocal differential conductance $G_{RL}(V) = dI(V) / dV$; here $I$ is the current flowing into the right lead, and $V$ is the bias applied to the left lead, see Fig.~\ref{fig:summary}(a).
At zero temperature, $G_{RL}(V)$ can be related to the transmission amplitudes of an electron across the wire in the following way:
\begin{equation}\label{eq:G}
G_{RL}(V) = G_Q \bigl(|S_{\rm ee}(E)|^2 - |S_{\rm he}(E)|^2\bigr)\bigr|_{E = eV}.
\end{equation}
We denoted the normal transmission amplitude of an electron from the left lead to the right by $S_{\rm ee}(E)$. In addition to being transmitted normally, an electron can undergo a particle-hole conversion in the transmission process; we denoted the respective amplitude by $S_{\rm he}(E)$. 
Equation~\eqref{eq:G} shows we need to solve the scattering problem to obtain $G_{RL}$.

\begin{figure}[t]
  \begin{center}
    \includegraphics[scale = 1]{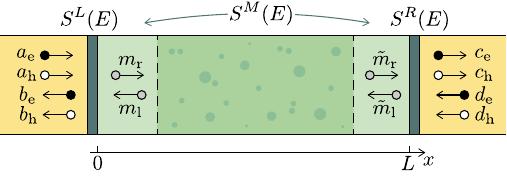}
    \caption{Setup of the scattering problem. Electron and hole waves in the leads are depicted with black and white circles, respectively. Gray circles represent Majorana modes in the wire. Scattering at the junctions between the wire (green) and the normal leads (yellow) is described by matrices  $S^{L}(E)$ and $S^{R}(E)$ [see Eqs.~\eqref{eq:Sl_def} and \eqref{eq:Sr_def}].  $S^M(E)$ is the scattering matrix of Majorana modes across the wire [see Eq.~\eqref{eq:Sm_def}].}
    \label{fig:waves}
  \end{center}
\end{figure}

A convenient way to find $S_{\rm ee}(E)$ and $S_{\rm he}(E)$ is to consider the scattering problems for the two lead-wire junctions and for the wire's bulk separately, and then match the {three} results with each other.
To make the individual scattering problems well-defined, we introduce a ``wave region'' next to each junction---i.e.,
a clean segment of the wire supporting propagating waves, see Fig.~\ref{fig:waves} [each such segment is described by Eq.~\eqref{eq:wire_c}].
This assumption is similar to a respective assumption made in, e.g., description of a mesoscopic SNS junction~\cite{beenakker1991}.
We denote the wave amplitudes in the leads and in the wave regions according to Fig.~\ref{fig:waves}.

Let us first consider scattering at the lead-wire junctions. 
Each junction can described by a $3\times 3$ scattering matrix relating incoming and outgoing waves.
We denote the scattering matrices by $S^{L}_{\alpha\beta}(E)$ and $S^{R}_{\alpha\beta}(E)$ for the left and right junction, respectively.
The indices $\alpha, \beta$ belong to a set $\{{\rm m, e, h}\}$, where ${\rm m}$ corresponds to a Majorana wave in the wire, and ${\rm e, h}$ correspond to the electron and hole waves in the lead.
Specifically, $S^{L}(E)$ relates the wave amplitudes according to
\begin{equation}\label{eq:Sl_def}
\begin{pmatrix}
    m_{\rm r}\\ b_{\rm e} \\ b_{\rm h}
    \end{pmatrix}
    =
    S^L(E)
    \begin{pmatrix}
     m_{\rm l}\\ a_{\rm e} \\ a_{\rm h}
    \end{pmatrix}.
\end{equation}
Similarly, $S^{R}(E)$ is defined by
\begin{equation}\label{eq:Sr_def}
    \begin{pmatrix}
    \tilde{m}_{\rm l}e^{-ik(E)L}\\ c_{\rm e} \\ c_{\rm h}
    \end{pmatrix}
    =
    S^R(E)
    \begin{pmatrix}
     \tilde{m}_{\rm r} e^{ik(E)L}\\ d_{\rm e} \\ d_{\rm h}
    \end{pmatrix}.
\end{equation}
Note the phase-factors in the latter expression. They describe the energy-dependent phases accumulated by the Majorana modes across the wire; the wave vector $k(E) = E / v$.

The scattering matrices $S^{L/R}(E)$ are determined by a number of non-universal properties of the junctions, and we will not attempt to find them microscopically. 
In fact, this is not required to establish the low-bias behavior of the conductance. 
We will demonstrate that, for the latter purpose, it is sufficient to constrain the form of $S^{L/R}(E)$ using the particle-hole transformation.
The particle-hole transformation relates an eigenstate with energy $E$ to the one with energy $-E$. It is implemented by the operator ${\cal P} = {\Theta}\,{\cal K}$. Here ${\cal K}$ is the complex conjugation; matrix $\Theta$ interchanges particle- and hole-components of the wavefunction in the lead, and is given by
\begin{equation}
    \Theta =
    \begin{pmatrix}
        1 & 0 & 0\\
        0 & 0 & 1\\
        0 & 1 & 0
    \end{pmatrix}
\end{equation}
(in the same basis as the one used in Eqs.~\eqref{eq:Sl_def} and \eqref{eq:Sr_def}).
Application of ${\cal P}$ to the scattering states stipulates the following relations for the $S$-matrices:
\begin{equation}\label{eq:ph_junctions}
    S^{R/L}(E) = \Theta [S^{R/L}(-E)]^\star \Theta.
\end{equation}

To demonstrate the significance of the above relation, let us apply it to the right junction at $E = 0$. From Eq.~\eqref{eq:ph_junctions}, one readily sees that 
$|S^R_{\rm em}(0)|^2 = |S^R_{\rm hm}(0)|^2$, i.e., at the Fermi level, the quasiparticle leaves the wire in an equal-weight superposition of electron and hole states. This means that $G_{RL}(0) = 0$ {\it regardless} of the wave dynamics in the wire.
Physically, the vanishing of the conductance stems from the neutral character of the Majorana modes. We will discuss further consequences of Eq.~\eqref{eq:ph_junctions} shortly.

Next, we consider the transmission of Majorana modes across the wire's bulk. The amplitudes $m_{\rm r, l}$ and $\tilde{m}_{\rm r, l}$ [see Fig.~\ref{fig:waves}] are linearly dependent. The dependence can be expressed in terms of the $2\times 2$ scattering matrix $S^M_{\alpha \beta}(E)$, where $\alpha, \beta \in \{\rm r, l\}$, and index ${\rm r/ l}$ corresponds to a right-/left-moving Majorana mode:
\begin{equation}\label{eq:Sm_def}
\begin{pmatrix}
\tilde{m}_{\rm r} e^{ik(E)L}\\m_{\rm l}
\end{pmatrix}
=
S^M(E)
\begin{pmatrix}
m_{\rm r} \\ \tilde{m}_{\rm l} e^{-ik(E)L}
\end{pmatrix}
\end{equation}
Obtaining $S^M(E)$ requires one to solve the Schr\"odinger equation corresponding to Hamiltonian \eqref{eq:H_wire}. We will do that in a number of cases below, see Secs.~\ref{sec:clean} and \ref{sec:disorder}. For now, we only note that the particle-hole transformation applied to the Majorana modes results in
\begin{equation}\label{eq:ph_majoranas}
    S^M(E) = [S^M(-E)]^\star.
\end{equation}

To find the two amplitudes entering the expression for $G_{RL}(V)$, cf.~Eq.~\eqref{eq:G}, we consider an electron impinging on the wire from the left lead. This amounts to setting $a_{\rm h} = d_{\rm e} = d_{\rm h} = 0$ and $a_{\rm e} = 1$ in Eqs.~\eqref{eq:Sl_def} and \eqref{eq:Sr_def} [cf.~Fig.~\ref{fig:waves}]. In this case, $c_{\rm e} \equiv S_{\rm ee}(E)$ and $c_{\rm h} \equiv S_{\rm he}(E)$.
Solving Eqs.~\eqref{eq:Sl_def},\eqref{eq:Sr_def}, and \eqref{eq:Sm_def} as a system of equations for $c_{\rm e}$, $c_{\rm h}$, $m_{\rm r, l}$, and $\tilde{m}_{\rm r,l}$, we find
\begin{widetext}
\begin{align}
\begin{pmatrix}
    S_{\rm ee}(E)\\
    S_{\rm he}(E) 
\end{pmatrix} =
\begin{pmatrix}
    S_{{\rm em}}^R(E)\\
    S_{{\rm hm}}^R(E)
\end{pmatrix}S_{{\rm me}}^L(E)\,
\frac{S_{\rm rr}^M(E)}{1 - (r_{\rm m}^R(E) S_{\rm rl}^M(E) - r_{\rm m}^L(E) S_{\rm lr}^M(E)) + r_{\rm m}^R(E)r_{\rm m}^L(E) \det S^M(E)},
\label{eq:she}
\end{align}
where we introduced $r^{R}_{\rm m}(E) \equiv S_{\rm mm}^R(E)$ and $r^{L}_{\rm m}(E) \equiv -S_{\rm mm}^L(E)$.
Substituting these expressions into Eq.~\eqref{eq:G}, we obtain for the conductance:
\begin{equation}\label{eq:G_RL_P}
    G_{RL}(V) / G_Q  = \bigl(T^R_{{\rm e m}}(E) - T^R_{{\rm h m}}(E)\bigr) {T_{{\rm me}}^L(E) Q(E)}\bigr|_{E = eV}.
\end{equation}
Here the transmission coefficients of the junctions are defined by $T_{\rm me}^L(E) \equiv |S_{\rm me}^L(E)|^2$ and $T^R_{i\rm m}(E) = |S_{i \rm m}^R(E)|^2$, with $i \in \{\rm e, h\}$. Function $Q(E) > 0$ is given by
\begin{equation}\label{eq:PE_general}
    Q(E) = \frac{|S_{\rm rr}^M(E)|^2 }{|1 - (r_{\rm m}^R(E) S_{\rm rl}^M(E) - r_{\rm m}^L(E) S_{\rm lr}^M(E)) + r_{\rm m}^R(E) r_{\rm m}^L(E) \det S^M(E)|^2};
\end{equation}
it describes propagation of the Majorana mode from one end of the wire to another [in particular, note that $Q(E) \propto |S_{\rm rr}^M(E)|^2$].
Equation \eqref{eq:G_RL_P} is the central  results of this section. It has a number of interesting properties which we now describe. 
\end{widetext}

First, it follows from Eq.~\eqref{eq:G_RL_P} that the nonlocal conductance is an {\it odd} function of bias at low $V$. 
To see this, let us examine Eq.~\eqref{eq:G_RL_P} term by term, starting with the expression in the round brackets. Using particle-hole symmetry \eqref{eq:ph_junctions}, one readily sees that $T_{\rm hm}^R(E) = T_{\rm em}^R(-E)$. Thus the expression in the brackets---i.e., $T^R_{{\rm e m}}(E) - T^R_{{\rm h m}}(E) = T^R_{{\rm e m}}(E) - T_{\rm em}^R(-E)$---is odd in $E$. On the contrary, application of Eqs.~\eqref{eq:ph_junctions} and \eqref{eq:ph_majoranas} shows that $Q(E) $ is an even function, $Q(E) = Q(-E)$. The last term, $T_{\rm me}^L(E)$, does not generally have any symmetry under $E \rightarrow -E$. However, in the vicinity of the Fermi level, it can be approximated by $T_{\rm me}^L(0)$. Summarizing, we conclude that $G_{RL}(V) = - G_{RL}(-V)$. 
In particular, $G_{RL}(0) = 0$, as was announced after Eq.~\eqref{eq:ph_junctions}.

{The found leading (in the wire length $L$) asymptote of $G_{RL}(V)$ is associated with the Majorana modes, the slowest-decaying modes in the wire. The other, high-energy modes are evanescent and decay faster. The evanescent modes may contribute to sub-leading in $L$ terms violating the antisymmetry of $G_{RL}(V)$ \cite{rect2018}.}

Second, Eq.~\eqref{eq:G_RL_P} demonstrates that the \textit{sign} of $G_{RL}(V)$ is determined only by the properties of the \textit{right} junction. Which sign is realized in a given device depends on the {details} of the junction. For example, if the right junction hosts a localized state with energy below the Fermi level, then the transmission of holes is favored; $G_{RL}(V) > 0$ at \textit{negative} biases. By contrast, if the junction hosts a level above $E = 0$, then the transmission of particles is favored, and $G_{RL}(V) > 0$ at \textit{positive} biases.

The observation that the sign of $G_{RL}(V)$ is determined \textit{only} by the properties of the right junction hints on a connection between $G_{RL}(V)$ and the local conductance $G_{RR}(V)$.
In fact, there indeed exists a relation between the two quantities.
In general, the unitarity of the $S$-matrix demands the odd-in-bias components of $G_{RR}(V)$ and $G_{RL}(V)$ to be the same \cite{flensberg2020},
\begin{equation}\label{eq:flensberg}
    G^{\rm odd}_{RR}(V) = G^{\rm odd}_{RL}(V),
\end{equation}
where $G^{\rm odd}_{ij}(V) = (G_{ij}(V) - G_{ij}(-V))/2$.
As we showed though, the nonlocal conductance is odd in $V$ at low biases. Therefore, Eq.~\eqref{eq:flensberg} reduces to $G^{\rm odd}_{RR}(V) = G_{RL}(V)$. In other words, the local conductance $G_{RR}(V)$ contains the full information about the low-bias behavior of the nonlocal conductance $G_{RL}(V)$ close to the critical point.

To conclude this section, we apply Eq.~\eqref{eq:G_RL_P} to bring $G_{RL}(V)$ to a convenient final form valid at low biases. Close to the Fermi level, $E = 0$, we can approximate $T^L_{\rm me}(E) \approx T^L_{\rm me}(0)$, as discussed above. We can also expand $T^R_{{\rm e}M}(E) \approx T^R_{{\rm e}M}(0) + \partial_E T^R_{{\rm e}M}(0) E$ and $T^R_{{\rm h}M}(E)\approx T^R_{{\rm e}M}(0) - \partial_E T^R_{{\rm e}M}(0) E$ [the form of the expansion for holes follows from the particle-hole symmetry, cf.~Eq.~\eqref{eq:ph_junctions}]. These approximations lead to $G_{RL}(V) = G_Q\cdot 2\partial_E T_{\rm em}^R(0) T_{\rm me}^L(0) Q(E)|_{E = eV}$. We further introduce the energy scale for the variation of $T_{\rm em}^R(E)$ with $E$,
\begin{equation}\label{eq:Sigma_R}
\Sigma_R = [2\partial_E T_{\rm em}^R(0) / T_{\rm em}^R(0)]^{-1}.
\end{equation}
In terms of $\Sigma_R$, we find
\begin{equation}\label{eq:G_RL_low_bias}
    G_{RL}(V) = G_Q\cdot\frac{E}{\Sigma_R}\cdot T_{\rm em}^R(0)T_{\rm me}^L(0)\cdot Q(E)\bigr|_{E = eV}.
\end{equation}
This equation is applicable both at the critical point and in its vicinity, both for clean wires and for disordered ones.
The information on closeness to the critical point and disorder is encoded into the function $Q(E)$. In the following sections, we will find $G_{RL}(V)$ for a number of practically interesting cases. 
We start by considering a clean wire tuned close to the critical point.

\section{Conductance of a clean wire\label{sec:clean}}
A wire short compared to the mean free path, $L \ll v \tau$, can be treated as a clean one ($\tau$ is the Drude mean free time).
In this section, we neglect the disorder and find $G_{RL}(V)$ for such a clean wire. 

\subsection{Conductance at the critical point\label{sec:clean_crit}}
To start with, we assume that the wire is tuned to the critical point. 
The quasiparticle propagation along a clean, critical wire is ballistic. 
Therefore, the scattering across the wire's bulk %matrix $S^M(E)$
boils down to %a factor describing 
the phase accumulation, 
$S^M(E) = \mathbbm{1}_{2 \times 2}\cdot e^{ik(E) L}$, where $k(E) = E / v$ and $\mathbbm{1}_{2 \times 2}$ is a $2\times 2$ identity matrix.
The phase factor varies with $E$ on a scale $\sim v / L$ set by the level spacing in the wire.

\begin{figure*}[t]
  \begin{center}        \includegraphics[scale = 1]{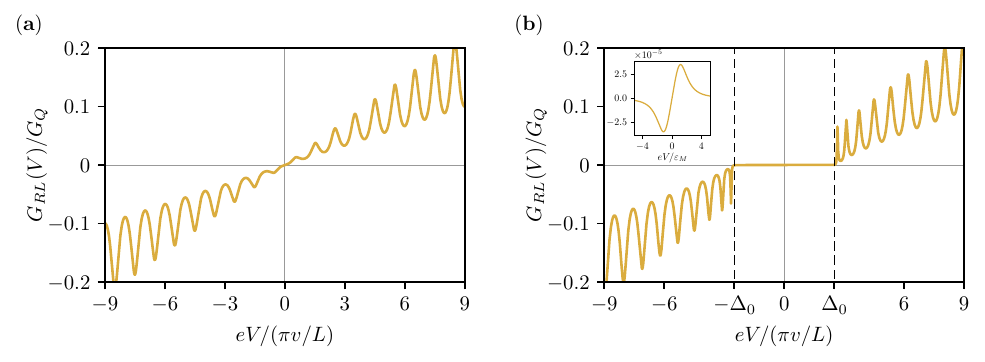}
    \caption{Nonlocal differential conductance $G_{RL}(V)$ of a clean wire at (panels {\bf a}) and away from (panel {\bf b}) the critical point at zero temperature. The curves are produced with the help of Eqs.~\eqref{eq:G_RL_low_bias}, \eqref{eq:clean_crit_PE}, \eqref{eq:FM}, and \eqref{eq:FM_above-the-gap}, in which we take $T_{\rm em}^L(0) = T^R_{\rm me}(0) = 0.4$ and
    $\Sigma_R = 50 \pi v / L $. In panel {\bf b}, we choose $\Delta_0 = 2.5 \pi v / L$; $\Delta_0 > 0$ corresponds to the topological phase. 
    The nonlocal conductance is an odd function of bias. At the critical point, it has a sequence of Fabry-P\'erot peaks with a spacing $e\delta V = \pi v / L$. Tuning the wire away from the critical point leads to opening of the ``transport'' in  gap of magnitude $|\Delta_0| \propto |B - B_{\rm c}|$.  
    Majorana zero modes are reflected in $G_{RL}(V)$ as subgap resonances (inset of panel {\bf b}). To make the resonances in $G_{RL}(V)$ well-resolved, in the inset we take $T_{\rm em}^L(0) = T^R_{\rm me}(0) = 10^{-3}$. In contrast to the resonance in the local conductance [see Fig.~\ref{fig:local} in Appenidx~\ref{sec:local}], the resonances in $G_{RL}(V)$ are exponentially small in the system size, cf.~Eqs.~\eqref{eq:splitting} and \eqref{eq:resonant}.}
    \label{fig:cond_clean}
  \end{center}
\end{figure*}

Using the expression for $S^M(E)$, we find that $G_{RL}(V)$ is given by Eq.~\eqref{eq:G_RL_P} with $Q(E)$ of the form:
\begin{equation}
    Q(E) = \frac{1}{{\bigl|1 + r_{\rm m}^R(E) r_{\rm m}^L(E) e^{\frac{2iEL}{v}}\bigr|^2}}.
\end{equation}
We will assume that the variation of the scattering amplitudes of the junctions with $E$ is slow on a scale $\sim v / L$.
In this case, to describe the low-bias behavior of $G_{RL}(V)$, we can replace $r_{\rm m}^{R/L}(E)$ by their values at the Fermi level, $r_{\rm m}^{R/L} \equiv r_{\rm m}^{R/L}(0)$.
Particle-hole symmetry \eqref{eq:ph_junctions} demands $r_{\rm m}^{R/L}$ to be real numbers \footnote{We note that amplitudes $r_{\rm m}^{j}$ are closely related to transmission coefficients $T_{i\rm m}^{j}(0)$ and  $T_{{\rm m}i}^{j}(0)$. Unitarity requires $r_{\rm m}^j = (1 - 2 T_{i\rm m}^j(0))^{1/2}$ and $T_{{\rm m}i}^{j}(0) = T_{i{\rm m}}^{j}(0)$.}. 
Then, we obtain
\begin{equation}\label{eq:clean_crit_PE}
    Q(E) = \frac{1}{{1 + [r_{\rm m}^R r_{\rm m}^L]^2 + 2 r_{\rm m}^R r_{\rm m}^L \cos(2EL/v)}}.
\end{equation}
The resulting $G_{RL}(V)$ is shown in Fig.~\ref{fig:cond_clean}(a). The oscillations with $E$ stem from the interference of paths with multiple reflections of the Majorana modes at the junctions. They are similar in origin to the transmission oscillations in a Fabry-P\'erot interferometer. 

If both junctions are transparent, $r^{R/L}_{\rm m} = 0$, then the differential conductance depends on bias linearly, $G_{RL} \propto V$ [cf.~Eqs.~\eqref{eq:G_RL_low_bias} and \eqref{eq:clean_crit_PE}]. This translates into a peculiar quadratic $I(V)$-characteristic:
\begin{equation}
    I(V) \propto V^2.
\end{equation}
The sign of the current is determined by the sign of $\Sigma_{R}$ in Eq.~\eqref{eq:Sigma_R}; it
is similar for $V > 0$ and $V < 0$. The latter property
allows one to use the system as a rectifier. 
{Such a rectifying behavior was first pointed out (and related to the particle-hole symmetry) in Ref.~\onlinecite{rect2018}.}

\subsection{Conductance away from the critical point\label{sec:clean_away}}
For a clean wire, tuning $B$ away from $B_{\rm c}$ leads to an opening of the spectral gap $E_{\rm gap} = |\Delta_0|$, where $\Delta_0 \propto B - B_{\rm c}$ [see Eq.~\eqref{eq:detuning} and related discussion]. 
In this section, we show how the presence of the gap is reflected in the conductance of the wire $G_{RL}(V)$.

The character of the quasiparticle propagation along the wire is different at subgap and above-the-gap energies. 
For $|E| > |\Delta_0|$, the quasiparticle propagates along the wire freely. 
In the opposite case, $|E| < |\Delta_0|$, the quasiparticle wave function decays exponentially along the wire.
This difference leads to a threshold at $V = |\Delta_0|/e$ in the bias-dependence of the conductance of a long wire; $G_{RL}(V)$ is exponentially small below the threshold and appreciable above it.

We now quantify this behavior starting with the general expression \eqref{eq:G_RL_P} for the conductance. 
%is given by a general expression \eqref{eq:G_RL_P}. 
To obtain function $Q(E)$ entering $G_{RL}(V)$ [Eq.~\eqref{eq:PE_general}], one needs to find the scattering matrix $S^M(E)$. This can be done by solving the Schr\"odinger equation defined by the Hamiltonian $H_{\rm wire} = H_{\rm wire, c} + \delta H_{\rm wire}$, see Eqs.~\eqref{eq:wire_c} and \eqref{eq:detuning}.
Focusing on subgap energies first, $|E| < |\Delta_0|$, we obtain after a straightforward calculation (see Appendix~\ref{sec:SM_derivation} for details):
\begin{align}\label{eq:Sm_solution}
    &\hspace{-0.1cm}S^M(E) =\notag\\
    &\hspace{-0.1cm}\frac{\begin{pmatrix}
    e^{L/ l}(1-e^{2i\eta}) & i\,\mathrm{sgn}\,\Delta_0\,(e^{{2L}/{l}} - 1)e^{i\eta}\\
    -i\,\mathrm{sgn}\,\Delta_0 (e^{{2L}/{l}}-1)e^{i\eta} & e^{L / l}(1-e^{2i\eta})
    \end{pmatrix}}{e^{2L/l} - e^{2i\eta}}.
\end{align}
Here $l \equiv l(E)$ is the length scale describing the decay of the wave function in the gapped region. It is given by
\begin{equation}\label{eq:xi_ind}
l(E) = \frac{v}{\sqrt{\Delta_0^2-E^2}}.
\end{equation}
We also introduced the phase factor in Eq.~\eqref{eq:Sm_solution}:
\begin{equation}\label{eq:eta}
    e^{i\eta(E)} = \frac{E}{|\Delta_0|} - i \sqrt{1-\frac{E^2}{\Delta_0^2}}.
\end{equation}
Notice that the backscattering phase in Eq.~\eqref{eq:Sm_solution} differs by $\pi$ for $\Delta_0 > 0$ and $\Delta_0 < 0$ (see the off-diagonal components of $S_M$).
The sign of $\Delta_0$ distinguishes the two sides of the topological transition. 
As discussed in Appendix~\ref{sec:gauge}, we fix the ``gauge'' of the Majorana fields in which the topological phase is realized at $\Delta_0 > 0$.

Next, we substitute Eq.~\eqref{eq:Sm_solution} into Eq.~\eqref{eq:PE_general}. Similarly to the derivation of Eq.~\eqref{eq:clean_crit_PE}, we dispense with the energy dependence of reflection amplitudes off the junctions, $r_{\rm m}^{R / L}(E) \rightarrow r_{\rm m}^{R / L}$. Then, we find at $|E| < |\Delta_0|$:
\begin{widetext}
\begin{equation}\label{eq:FM}
Q(|E| < |\Delta_0|) = \frac{\bigl(\Delta_0^2 - E^2 \bigr)}{\Bigl[\sqrt{\Delta_0^2-E^2} \cosh\bigl[\frac{L}{l(E)}\bigr]\bigl(1+r_{\rm m}^Rr_{\rm m}^L\bigr) - \Delta_0 \bigl(r_{\rm m}^R + r_{\rm m}^L\bigr) \sinh\bigl[\frac{L}{l(E)}\bigr]\Bigr]^2 + E^2 \sinh^2\bigl[\frac{L}{l(E)}\bigr](1-r_{\rm m}^R r_{\rm m}^L)^2}.
\end{equation}
To obtain $Q(E)$  at energies above the gap, we perform an analytic continuation in Eq.~\eqref{eq:Sm_solution} from $|E| < |\Delta_0|$ to $|E| > |\Delta_0|$. In this way, we find
\begin{equation}\label{eq:FM_above-the-gap}
Q(|E| > |\Delta_0|) = \frac{\bigl(E^2 - \Delta_0^2 \bigr)}{\Bigl[\sqrt{E^2-\Delta_0^2} \cos[\kappa(E)L]\bigl(1+r_{\rm m}^Rr_{\rm m}^L\bigr) - \Delta_0 \bigl(r_{\rm m}^R + r_{\rm m}^L\bigr) \sin[\kappa(E)L]\Bigr]^2 + E^2 \sin^2[\kappa(E) L](1-r_{\rm m}^R r_{\rm m}^L)^2},
\end{equation}
\end{widetext}
where $\kappa(E) = \sqrt{E^2 - \Delta_0^2}/v$. In the limit $\Delta_0 \rightarrow 0$, the latter expression reproduces the result found at the critical point, Eq.~\eqref{eq:clean_crit_PE}. A notable feature of Eqs.~\eqref{eq:FM} and \eqref{eq:FM_above-the-gap} is that $Q(E)$ vanishes at the gap edge, $E = |\Delta_0|$. 

Equation \eqref{eq:FM} shows that the non-local conductance is exponentially suppressed at subgap energies. We find $G_{RL}(V) \propto Q(eV) \propto \exp[- 2 L / l(eV)]$ for a long wire, $L \gg l(eV)$.
The exponential suppression ceases at the threshold bias $|V| =  |\Delta_0| / e$. 
At higher biases, $|V| \gtrsim |\Delta_0| / e$, the conductance experiences Fabry-P\'erot oscillations with $V$, see Fig.~\ref{fig:cond_clean}(b). 

To summarize, {\it in a clean wire}, one can directly access the spectral gap $E_{\rm gap} = |\Delta_0|$ by finding the threshold in $G_{RL}(V)$.
In particular, the measurement of $G_{RL}(V)$ can be used to pinpoint the position---as a function of field $B$ or other control parameters such as the chemical potential $\mu$---of the topological phase transition. The critical point of the transition is the one at which the gap closes, $\Delta_0 = 0$.

\subsection{Role of Majorana zero modes\label{sec:majoranas}}
A defining feature of the topological phase ($\Delta_0 > 0$) is the presence of MZMs localized at the wire's ends. It is well-known that MZMs lead to a zero-bias peak in {\it local} conductance [see supplementary Fig.~\ref{fig:local}(b)].
How are the zero modes reflected in the {\it nonlocal} conductance?
To answer this question, we consider a wire of length $L \gg v / \Delta_0$ that is connected to the leads by the tunnel (i.e., weakly transparent) junctions. For simplicity, we assume that the junctions are similar and take 
\begin{equation}\label{eq:Ts}
    T_{{\rm e m}}^{R}(0) = T_{{\rm m e}}^{L}(0) = T / 2 \ll 1.
\end{equation}
The unitarity requires that the reflection amplitudes are given by $r^R_{\rm m} = r^L_{\rm m} = \sqrt{1 - T}\approx 1 - T/2$. 
We will characterize the junctions by the coupling rate
\begin{equation}\label{eq:Gamma}
\Gamma \equiv \Delta_0 T / 2 \ll \Delta_0,
\end{equation}
which has units of energy.

A finite size of the wire $L$ leads to the hybridization of MZMs. 
The energy of the hybridized level is separated from the Fermi level by $\varepsilon_M$, which is exponentially small in $L$ for a long wire, $L \gg v / \Delta_0$:
\begin{equation}\label{eq:splitting}
    \varepsilon_M = 2\Delta_0 e^{-L\Delta_0 / v} \ll \Delta_0. 
\end{equation}
It is instructive to express the conductance in terms of $\varepsilon_M$ and $\Gamma = \Delta_0 T / 2$.  Focusing on small biases $|V| \ll \Delta_0 / e$ and using Eqs.~\eqref{eq:G_RL_low_bias} and \eqref{eq:FM}, we find for $G_{RL}(V)$:
\begin{equation}\label{eq:resonant}
    G_{RL}(V) = G_Q \cdot \frac{\varepsilon_M}{4 \Sigma_R}\cdot \sum_{s = \pm} \frac{s\,\Gamma^2}{(eV-s\,\varepsilon_M)^2 + \Gamma^2}
\end{equation}
[we recall that the energy scale $\Sigma_R$ here characterizes the variation of the right junction's transparency with energy, see Eq.~\eqref{eq:Sigma_R}].
The conductance has a structure of two superimposed Lorentzians of opposite signs that are centered at $V = \pm \varepsilon_M / e$ [see the inset of Fig.~\ref{fig:cond_clean}(b)]. 
The Lorentzians do not overlap provided the wire's coupling to the leads is sufficiently weak, $\Gamma \ll \varepsilon_M$; we focus on this limit in the rest of the section.

Equation \eqref{eq:resonant} demonstrates that MZMs lead to the resonant transmission across the sample.
In fact, the transmission probabilities in the particle and hole channels are both of the order of unity at resonances: $|S_{\rm ee}(\varepsilon_M)|^2 \approx |S_{\rm he}(\varepsilon_M)|^2 \approx 1 / 4$, as follows from Eqs.~\eqref{eq:she} and \eqref{eq:FM}.
However, it is their {\it difference} that determines the conductance, see Eq.~\eqref{eq:G}.
The difference is suppressed by a factor $\varepsilon_M / \Sigma_R$ (which is exponentially small in $L$), resulting in the respective suppression of $G_{RL}(V)$, see Eq.~\eqref{eq:resonant}.

Although the conductance is exponentially small at resonances, it still vastly exceeds $G_{RL}(V)$ in the trivial phase evaluated at the same bias. Indeed, in the latter case, we find with the help of Eq.~\eqref{eq:FM}:
\begin{equation}\label{eq:triv}
G_{RL}(V) \approx  G_Q\cdot \frac{eV}{\Sigma_R} \cdot \Bigl[\frac{2\Gamma}{\Delta_0}\Bigr]^2\cdot e^{-2L|\Delta_{0}|/v}.
\end{equation}
By comparing Eqs.~\eqref{eq:resonant} and \eqref{eq:triv} at $V = \varepsilon_M / e$ under the above-made assumption $\varepsilon_M \gg \Gamma$, we see that $G_{RL}(V)$ in the topological phase exceeds that in the trivial phase by a factor $\sim [\Delta_0/\Gamma]^2 \cdot e^{2L \Delta_0/v} \gg 1$.

\subsubsection*{Signature of MZMs in the shot noise}
The smallness of the MZM-induced resonance in the conductance stems from an almost complete cancellation between $|S_{\rm ee}(E)|^2$ and $|S_{\rm he}(E)|^2$. No such cancellation occurs in the nonlocal shot noise ${\cal S}_{RL}(V)$, which is another quantity directly accessible via an electron transport measurement.
In fact, MZMs lead to the quantized resonances in the differential noise $\partial {\cal S}_{RL}/\partial V$, as we will now show.

A counterpart of Eq.~\eqref{eq:G_RL_low_bias} for the differential noise is
\begin{equation}\label{eq:diff_noise}
\frac{\partial {\cal S}_{RL}(V)}{\partial V} = \frac{e^3}{\pi}{T_{{\rm me}}^L(0) T^R_{{\rm em}}(0) Q(eV)}.
\end{equation}
This is a general result for ${\cal S}_{RL}(V)$ applicable near the critical point, see Appendix \ref{sec:diff_noise_derivation} for the derivation.

Let us now apply Eq.~\eqref{eq:diff_noise} to demonstrate how MZMs manifest in the shot noise. Assuming that the resonances are narrow, $\Gamma \ll \varepsilon_M$, and using Eqs.~\eqref{eq:FM} and \eqref{eq:Ts}--\eqref{eq:splitting}, we find:
\begin{equation}\label{eq:noise_resonant}
    \frac{\partial {\cal S}_{RL}(V)}{\partial V} = \frac{e^3}{4\pi} \sum_{s=\pm} \frac{\Gamma^2}{(eV-s\,\varepsilon_M)^2 + \Gamma^2}.
\end{equation}
Remarkably, the differential noise takes a quantized value $e^3/4\pi$ at resonance, $V = \pm \varepsilon_M / e$. This is in a striking contrast to the behavior of the conductance, for which we found $G_{RL}(\pm \varepsilon_M / e) \ll G_Q$. We note in passing that the quantization of shot noise breaks down in the limit of strongly broadened resonances, $\Gamma \gtrsim \varepsilon_M$. {Then, a two-peak structure in $\partial_V S_{RL}(V)$ of Eq.~\eqref{eq:noise_resonant} is replaced by a single zero-bias peak of width $\sim \Gamma$ and height $\sim e^3 \varepsilon_M^2 / (\pi \Gamma^2)$.} 

It is instructive to characterize the shot noise by the Fano factor $F = {\cal S}_{RL} / eI$. We will focus on biases satisfying $\varepsilon_M \ll eV \ll |\Delta_0|$ and {$\Gamma \ll \varepsilon_M$}. For such biases, we obtain from Eqs.~\eqref{eq:resonant} and \eqref{eq:noise_resonant}:
\begin{equation}
    I = { \frac{e \Gamma}{8} }\frac{\varepsilon_M}{\Sigma_R},\quad\quad {\cal S}_{RL} = \frac{e^2\Gamma}{4}.
\end{equation}
Consequently, the Fano factor is given by
\begin{equation}\label{eq:Fano}
    F = \frac{2\Sigma_R}{\varepsilon_M}\propto \exp (L\Delta_0 / v).
\end{equation}
It is \textit{exponentially large} in the system size $L$. We note, however, that this largeness  merely reflects the smallness of the average current, and does not imply the bunching of charges (as in, e.g., Refs.~\onlinecite{vayrynen2017, kurilovich2019}). 

\section{Nonlocal conductance of a disordered wire\label{sec:disorder}}
So far, we have neglected the disorder and obtained the nonlocal conductance $G_{RL}(V)$ of a clean wire.
The clean-wire limit is relevant if the length of the wire $L$ is small compared to the mean free path $v \tau$. 
In this section, we find instead the conductance of a long wire, $L \gg v \tau$, for which disorder cannot be neglected.

In the presence of disorder, all states in the wire are localized.
This makes $G_{RL}(V)$ exponentially small in the wire's length $L$.
It is thus convenient to characterize the nonlocal transport by the conductance logarithm $\ln |G_{RL}(V) / G_Q|$ \footnote{We introduce the absolute value under the logarithm to account for the fact that $G_{RL}(V)$ can be of either sign.}. 
In contrast to $G_{RL}(V)$, this quantity is self-averaging (i.e., its value in a given device is representative of the ensemble of devices). We can represent
\begin{equation}\label{eq:exp_small}
\langle\ln |G_{RL}(V) / G_Q| \rangle = - \frac{2L}{l(E)}\Bigr|_{E = eV} + \dots\,.
\end{equation}
Here $l(E)$ has a meaning of the localization length of a quasiparticle with energy $E$, and $\dots$ encode contributions independent of the wire's length coming from the first two factors in Eq.~\eqref{eq:G_RL_P}. 
Below, we study the dependence of the localization length $l(E)$ on energy at the critical point and in its vicinity.

In the low-energy theory, the disorder is described by a term \eqref{eq:disorder} in the Hamiltonian.
We will assume that the field $\delta \Delta (x)$ in it is a Gaussian random quantity with short-ranged correlations,
\begin{equation}\label{eq:correlator}
    \langle \delta \Delta(x) \delta \Delta(x^\prime) \rangle = \frac{v}{\tau}\delta(x - x^\prime).
\end{equation}
This assumption is justified provided (i) the individual impurities in the wire are weak (i.e., it is many impurities that determine the scattering cross-section, as opposed to a single or a few ``strong'' impurities), and (ii)~the quasiparticle wavelength in the wire exceeds the correlation radius of the random potential, $r_{\rm c} \ll v / E$.
We parameterized the correlator in Eq.~\eqref{eq:correlator} 
by a time scale $\tau$. As we will see shortly, this time scale is directly related to the localization length at relatively high energies [see, e.g.,   Eq.~\eqref{eq:high_E_l_E} below].

Localization properties of the Majorana modes in the wire---including their localization length $l(E)$---are encoded in scattering matrix $S^M(E)$ defined in Eq.~\eqref{eq:Sm_def}. We now describe how $S^M(E)$ can be obtained.
A convenient approach is to reduce the problem of finding $S^M(E)$ to a solution of a system of differential equations.
A way to do it is to break the wire into a series of short elements (labelled by their position $x$), and then track how the scattering matrix $S^M(E, x)$ evolves as  the elements are joined together.
The scattering off a single short element can be described with the help of the Born approximation.
The addition of this element to the wire leads to a small, random change in $S^M(E,x)$. One can describe such changes with an analog of the Langevin equation.

To implement the outlined procedure, we first introduce a convenient parameterization for the scattering matrix:
\begin{equation}\label{eq:S_param}
S^M(E,x) =
\begin{pmatrix}
{e^{i\varphi_{\rm fwd}}}/{\cosh \lambda} & -\tanh \lambda\,e^{i\vartheta}\\
\tanh \lambda\,e^{i\beta - i \vartheta} & {e^{i \beta - i\varphi_{\rm fwd}}}/{\cosh \lambda}
\end{pmatrix}.
\end{equation}
Here $\vartheta \equiv \vartheta(E, x)$ and $\varphi_{\rm fwd}\equiv \varphi_{\rm fwd}(E,x)$  are the backscattering and forward scattering phases, respectively. Phase $\beta \equiv \beta (E,x)$ distinguishes the transmission amplitudes across the wire in the two directions.
Finally, $\lambda \equiv \lambda(E, x)$ is the Lyapunov exponent. The latter parameter controls the magnitude of scattering amplitudes, and is directly related to the localization length:
\begin{equation}\label{eq:l_E_lambda}
l(E) = \lim_{L\rightarrow \infty} [\langle |\lambda(E, L)|\rangle / L]^{-1}.
\end{equation}
Thus, we focus on obtaining $\lambda(E, L)$ for the purpose of finding $l(E)$. 

By considering the addition of a single short element, we find that the dynamics of $\lambda$ and $\vartheta$ with $x$ separate from the dynamics of $\varphi_{\rm fwd}$ and $\beta$. {In view of Eq.~\eqref{eq:l_E_lambda}}, only the former pair of variables is relevant for $l(E)$. We obtain the following system for the evolution of $\lambda$ and $\vartheta$ (see Appendix~\ref{sec:derivation_evol} for details):
\begin{subequations}\label{eq:dynamics_finite_E}
\begin{align}
    \frac{d\lambda}{dx} &= -\frac{\Delta(x)}{v} \cos \vartheta,\label{eq:lambda_evol}\\
    \frac{d\vartheta}{dx} &= \frac{2E}{v} + \frac{2\Delta(x)}{v} \sin \vartheta \coth 2\lambda.\label{eq:backscattering}
\end{align}
\end{subequations}
Here  $\Delta (x) = \Delta_0 + \delta \Delta (x)$ incorporates both the detuning from the critical point $\Delta_0 \propto B - B_{\rm c}$ and the disorder $\delta \Delta(x)$. The initial condition for the $S$-matrix is $S^M(E, 0) = \mathbbm{1}_{2\times 2}$ which translates into $\lambda (E , 0) = 0$. Because the off-diagonal component of the $S$-matrix vanishes at $x=0$ [cf.~Eq.~\eqref{eq:S_param}], we are free to set the initial condition for $\vartheta$; we fix $\vartheta(E , 0) = 0$.
{In passing, we note that at zero bias ($E = 0$) Eq.~\eqref{eq:backscattering} allows for a solution $\vartheta = 0$ or $\pi$. 
Substitution of this solution into Eq.~\eqref{eq:lambda_evol} yields $\lambda(x) = \mp \int_0^{x}\Delta(x^\prime) dx^\prime / v$. This shows that $\langle\lambda(x)\rangle$ grows linearly with $x$, while ${\rm std}[\lambda(x)] \propto \sqrt{x}$, reinforcing the notion of self-averaging of the conductance logarithm. }

We now apply Eqs.~\eqref{eq:l_E_lambda} and \eqref{eq:dynamics_finite_E} to derive a convenient general expression for the localization length $l(E)$.
To start with, we note that $\lambda \gg 1$ for a long wire \footnote{We assume hereafter that $\lambda > 0$. This condition can be always guaranteed by a phase shift, $\vartheta \rightarrow \vartheta + \pi$.  
}; thus, one can approximate $\coth 2\lambda \approx 1$ in Eq.~\eqref{eq:backscattering}. After making this approximation, we see that the evolution of the backscattering phase $\vartheta$ decouples from that of $\lambda$. On the contrary, $\lambda$ directly follows the dynamics of $\vartheta$ according to Eq.~\eqref{eq:lambda_evol}. One can thus find $l(E)$ in two steps. The first step is to determine statistical properties of $\vartheta$. Then, these properties can be used to evaluate $\langle \lambda \rangle$ and, therefore,~$l(E)$.

Statistical properties of the backscattering phase are encoded in distribution function $P(\vartheta|x)$. To find $P(\vartheta|x)$, we derive a Fokker-Planck equation governing the evolution of this distribution with $x$ (see, e.g., Ref.~\onlinecite{yamamoto2021}):
\begin{equation}\label{eq:fokker-planck_st}
    v\tau\frac{\partial P}{\partial x} = 2\frac{\partial}{\partial \vartheta} \bigl(\sin \vartheta \frac{\partial}{\partial \vartheta} (\sin \vartheta P)\bigr) - 2\tau\frac{\partial}{\partial \vartheta}\bigl([E + \Delta_0 \sin\vartheta]P\bigr).
\end{equation}
The first term on the right hand side describes the diffusion of the backscattering phase due to the disorder. The second term describes the drift of the phase, which stems from a finite energy $E$ and the detuning from the critical point $\Delta_0$. We note that the form of the drift term is similar to that for a particle moving in the ``washboard'' potential $U_{\rm eff}(\vartheta) = - E \vartheta + \Delta_0 (\cos \vartheta - 1)$.
For a long wire, the distribution function reaches a steady state $P(\vartheta)$, in which the drift and the diffusion are balanced with each other.

Next, we relate $\langle \lambda \rangle$ to steady state distribution $P(\vartheta)$.
First, by averaging both sides of Eq.~\eqref{eq:backscattering}, we obtain $d\langle \lambda\rangle / dx = -\langle (\Delta(x)/v)\cos\vartheta\rangle$. The average on the right hand side can be expressed in terms of the harmonics of $P(\vartheta)$ using an approach outlined in Ref.~\onlinecite{ovchinnikov1980}. We find for $l^{-1}(E)\equiv \lim_{L\rightarrow\infty} \langle |\lambda(E, L)|\rangle / L$:
\begin{equation}\label{eq:loc_length}
\frac{1}{l(E)} = -\frac{\Delta_0}{v} \langle \cos \vartheta \rangle + \frac{1}{v \tau} \langle \sin^2 \vartheta \rangle, 
\end{equation}
where $\langle \dots \rangle = \int_{-\pi}^{\pi} d\vartheta \, (\dots) P(\vartheta)$.

We now use the obtained equations \eqref{eq:fokker-planck_st} and \eqref{eq:loc_length} to analyze the behavior of $l(E)$ at the critical point [Sec.~\ref{sec:disorder_crit}] and away from it [Secs.~\ref{sec:disorder_away} and \ref{sec:disorder_strong}]. {It follows from the properties of $Q(E)$ that $l(E)$ is an even function of $E$ [see Eq.~\eqref{eq:PE_general} and the discussion after it]. Keeping this in mind, below we assume $E > 0$.}

\subsection{Localization at the critical point\label{sec:disorder_crit}}
Here we consider a disordered wire tuned to the critical point, $\Delta_0 = 0$.
The critical point of the ``dirty'' wire belongs to the infinite-randomness universality class \cite{fisher1994, motrunich2001}. We show how the peculiar character of this critical state is reflected in the localization properties of the wire. 

At the critical point, $\Delta_0 = 0$, expression~\eqref{eq:loc_length} for $l(E)$ acquires the form
\begin{equation}\label{eq:loc_length_crit}
\frac{1}{l(E)} = \frac{1}{v\tau} \langle \sin^2 \vartheta \rangle,  
\end{equation}
where the average is evaluated with respect to the steady state distribution function $P(\vartheta)$ of the Fokker-Planck equation
\begin{equation}\label{eq:FP_crit}
    v\tau \frac{\partial P}{\partial x} = 2\frac{\partial}{\partial \vartheta} \bigl(\sin \vartheta \frac{\partial}{\partial \vartheta} (\sin \vartheta P)\bigr) - 2E\tau\frac{\partial}{\partial \vartheta}P = 0.
\end{equation}

Let us first apply these equations to find the localization length at relatively high energies, $E \gg 1/\tau$. In this case, the main term in Eq.~\eqref{eq:FP_crit} is the drift term. By discarding a subleading diffusion term, we find a uniform distribution function $P(\vartheta) = 1/2\pi$. Using it to compute the average in Eq.~\eqref{eq:loc_length_crit}, we obtain $\langle \sin^2 \vartheta \rangle = 1/2$ and thus
\begin{equation}\label{eq:high_E_l_E}
    l(E\gg 1 / \tau) = 2 v \tau.
\end{equation}
This shows that at high energies (which we nonetheless assume to be much smaller than the gap in the parent superconductor), the localization length is independent of energy and directly proportional to the Drude time $\tau$. 

It is instructive to assess how $l(E)$ deviates from Eq.~\eqref{eq:high_E_l_E} as the energy is lowered. This can be done by treating the first term in Eq.~\eqref{eq:FP_crit} as a perturbation. Using the perturbation theory to find $P(\vartheta)$ to the second order in $1/E\tau \ll 1$, and computing the average in Eq.~\eqref{eq:loc_length_crit}, we obtain
\begin{equation}\label{eq:crit_pert}
    l(E) \approx 2v\tau \Bigl(1 + \frac{1}{4E^2 \tau^2}\Bigr).
\end{equation}
Interestingly, the localization length \textit{increases} upon decreasing $E$. This is a precursor of the unusual low-energy behavior of $l(E)$ which we now turn to. 

At $E \sim 1 / \tau$, the two terms in the brackets in Eq.~\eqref{eq:crit_pert} become of the same order. This signals the breakdown of the perturbation theory. At lower energies, $E \lesssim 1 / \tau$, the diffusion and the drift terms in Fokker-Planck equation \eqref{eq:FP_crit} have to be treated on the same footing.
To do that, it is convenient to introduce a new variable $w$, which would bring the diffusion term to a standard form $\partial^2 P / \partial w^2$.
The respective change of variables reads~\footnote{The presented change of variables is applicable in the interval $\vartheta \in [0, \pi]$.
For $\vartheta \in [-\pi, 0]$, the change of variables reads $w = \ln \tan [\vartheta + \pi] / 2$.}
\begin{equation}
    w = \ln \tan (\vartheta / 2). 
\end{equation}
In terms of the $w$-variable, the expression for $l(E)$ acquires the form:
\begin{align}\label{eq:loc_length_w}
    \frac{1}{l(E)} = \frac{1}{v \tau} \Bigl\langle\frac{1}{\cosh^2 w}\Bigr\rangle.
\end{align}
The averaging here is carried over the steady state distribution function $P(w) \equiv P(\vartheta) d\vartheta / dw$ of the Fokker-Planck equation
\begin{equation}\label{eq:FP_w}
    v\tau \frac{\partial P}{\partial x} = 2\frac{\partial^2 P}{\partial w^2} - 2E \tau\frac{\partial}{\partial w} (\cosh w P).
\end{equation}
The right hand side of this equation is similar to the diffusion kernel for a conventional Brownian particle moving in an external field. The potential of the field is $U_{\rm eff}(w) = - 2E\tau \sinh w$.  

{Equations~\eqref{eq:loc_length_w} and \eqref{eq:FP_w} allow one to find $l(E)$ numerically in the entire energy range (see Appendix~\ref{sec:general_l_E} and Fig.~\ref{fig:summary}(e), and analytically in the limit $E\ll 1/\tau$.} In the remainder of the section, we focus on the {latter} low-energy regime.
%, $E \ll 1 / \tau$. 
In it, the effective potential acquires a peculiar structure: $U_{\rm eff}(w)$ rises quickly at $w < -\ln (1 / E\tau)$, has a {broad} nearly equipotential region $w \in [-\ln(1 / E\tau), \ln (1 / E\tau)]$, and then drops sharply at $w > \ln (1 / E\tau)$.
To find the steady-state distribution function $P(w)$, it is useful to visualize the motion of the Brownian $w$-``particle'' in such a potential. 
Let us assume that the particle is initialized at $w = -\infty$ (i.e., $\vartheta = 0$).
First, the particle is rapidly swept by the external field to the beginning of the equipotential region, $w = -\ln(1 / E\tau)$. 
%[in ``time'' $\Delta x \sim l_0$]. 
From there, it moves diffusively until it reaches a ``cliff'', $w  = \ln(1 / E\tau)$.
After that, the external field quickly brings the particle to $w = +\infty$. When this happens, $w$ resets to $-\infty$, as required by the continuity of the evolution of $\vartheta$. Then the cycle repeats.
The steady-state distribution function $P(w)$ describing the cyclic motion should satisfy $\partial^2 P /\partial w^2 = 0$ in the equipotential region $[-\ln (1 / E\tau), \ln (1 / E\tau)]$, and vanish at the right end of the region, $P(\ln (1 / E\tau)) = 0$. These conditions result in
\begin{equation}\label{eq:distribution_crit_low}
P(w) = {\Theta(\ln (1/E\tau) - |w|)} \frac{1 - \frac{w}{\ln(1/E\tau)}}{2\ln (1/E\tau)}.
\end{equation}
The linear dependence on $w$ reflects the fact that $P(w)$ is characterized by a constant probability current.

We now use this distribution function to find $l(E)$. Substituting Eq.~\eqref{eq:distribution_crit_low} into Eq.~\eqref{eq:loc_length_w}, we obtain
\begin{equation}\label{eq:l_E_crit}
l(E) = v\tau \ln \Bigl(\frac{1}{E \tau}\Bigr).
\end{equation}
This result is valid up to the corrections $\sim v\tau$, which are relatively small at low energies, $\ln (1/E \tau) \gg 1$.

\begin{figure}[t]
  \begin{center}
    \includegraphics[scale = 1]{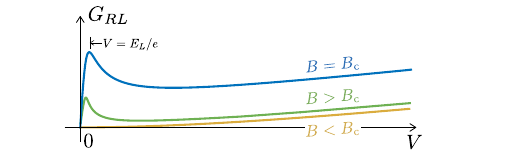}
     \caption{
     {Schematic plot of the bias dependence of the nonlocal conductance at and away from the critical point. At $B = B_{\rm c}$,  $G_{RL}(V)$ has a low-bias peak at $V = E_L / e$ [see Eq.~\eqref{eq:EL_scale}]. Both the position and the height of the peak scale exponentially with the length of the wire, $\propto \exp(-c_L L / v\tau)$. 
     Tuning magnetic field away from the the critical point reduces the conductance magnitude. On the trivial side of the transition, the peak in $G_{RL}(V)$ disappears entirely. On the topological side, the peak remains but its height is reduced compared to $B = B_{\rm c}$ value. Outside the critical region, the peak height is $\propto \exp(-L \Delta_0 / v)$, see Eq.~\eqref{eq:splitting}. The conductance at negative biases satisfies $G_{RL}(-V) = - G_{RL}(V)$.} 
     }
    \label{fig:close_away}
  \end{center}
\end{figure}

Remarkably, the localization length diverges at $E \rightarrow 0$.
The divergence indicates that---despite the disorder---the quasiparticle transmission across the critical wire becomes perfect at the Fermi level, $S^M_{\rm rr}(0) = 1$. 
The latter property was previously pointed out in Ref.~\onlinecite{akhmerov2011}.
It was proposed there as a signature of the topological phase transition that is manifested in the quantized thermal conductance and shot noise.
Measurement of the nonlocal conductance is easier to implement. The divergence of localization length at the Fermi level results in a low-bias peak in  $G_{RL}(V)$. Since $G_{RL}(0) = 0$ [cf.~Eq.~\eqref{eq:G_RL_low_bias}], the peak is displaced from zero to a bias $V = E_L / e$, see Fig.~\ref{fig:close_away}. We can estimate $E_L$ as an energy scale at which the transmission coefficient takes values $\sim 1$; this amounts to setting $l(E_L) \sim L$ in Eq.~\eqref{eq:l_E_crit}. From this condition we find
\begin{equation}\label{eq:EL_scale}
    E_L \sim \frac{1}{\tau}\exp\bigl(- c_L L / v \tau\bigr),
\end{equation}
with number $c_L \sim 1$. Because the conductance is odd in $V$, in addition to a peak at positive bias $V = E_L / e$, $G_{RL}(V)$ has a peak at $V = -E_L / e$ of opposite sign.
We note that the linear in $V$ term in Eq.~\eqref{eq:G_RL_low_bias} makes the (typical) peak magnitude exponentially small in $L$, $G_{RL}(E_L / e) \propto E_L \propto \exp (-c_L L / v\tau)$.

Energy scale $E_L$ also sets a constraint on temperature and bias at which the quantized responses predicted in Ref.~\onlinecite{akhmerov2011} can be observed.
Specifically, observation of the quantized thermal conductance requires $T \ll E_L$. Quantization of the shot noise occurs at $T \ll eV \ll E_L$.
 
The described logarithmic behavior of the localization length $l(E)$ [Eq.~\eqref{eq:l_E_crit}] can be directly linked to the low-energy behavior of the density of states (DOS) $\nu(E)$.
Due to the disorder, the DOS of a critical wire is singular at the Fermi level \cite{motrunich2001}:
\begin{equation}\label{eq:Dyson}
    \nu(E) \propto \frac{1}{E \ln^3\bigl(\frac{1}{E\tau}\bigr)},
\end{equation}
see Fig.~\ref{fig:summary}(e). This Dyson singularity \cite{dyson1953} is one of the key features of the infinite-randomness critical state.
According to Thouless \cite{thouless1972}, localization length $l(E)$ and DOS $\nu(E)$ are related to each other via an integral equation.
In the presence of particle-hole symmetry, the Thouless relation can be cast in the following form:
\begin{equation}\label{eq:Thouless}
    \frac{1}{l(E)} = \frac{1}{l(0)}+\int_0^{\infty} \nu(E^\prime) \ln\Bigl[\frac{|E^2 - E^{\prime 2}|}{E^{\prime 2}}\Bigr] dE^\prime.
\end{equation}
Substituting Eq.~\eqref{eq:Dyson} into this expression, 
%into Eq.~\eqref{eq:Thouless} 
evaluating the integral over $E^\prime$, and using $l(0) = \infty$, we recover Eq.~\eqref{eq:l_E_crit}.

The localization length characterizes the average value of Lyapunov exponent $\lambda(E, L)$ of a long wire, cf.~Eq.~\eqref{eq:l_E_lambda}.
In fact, in the low-energy limit, $E \ll 1/\tau$, one can go beyond the average value and evaluate the full distribution function $\Pi_0(\lambda)$. As we show in Appendix~\ref{sec:lyap_disr_appendix}, distribution $\Pi_0(\lambda)$ is Gaussian; its mean and variance are given by
\begin{equation}\label{eq:distr_props}
    \langle\lambda\rangle = \frac{L} {v\tau \ln\bigl(\frac{1}{E\tau}\bigr)}, \quad\quad \langle\hspace{-0.05cm}\langle \lambda^2 \rangle\hspace{-0.05cm}\rangle = \frac{L}{3v\tau}.
\end{equation}

Because the Lyapunov exponent is distributed normally, the respective distribution of the transmission coefficient ${\cal T} \equiv |S_{\rm rr}^M|^2 \approx 4 e^{-2\lambda}$ [cf.~Eq.~\eqref{eq:S_param}] is {\it log-normal}. This means that ${\cal T}$ is not a self-averaging quantity, with giant mesoscopic fluctuations. The \textit{typical} value of the transmission coefficient, ${\cal T}_{\rm typ} = 4 e^{-2\langle \lambda \rangle}$, is
\begin{equation}\label{eq:T_typ}
    {\cal T}_{\rm typ}(E) \sim \exp\Bigl[- \frac{2L}{v\tau\ln\bigl(\frac{1}{E\tau}\bigr)}\Bigr].
\end{equation}
{A similar scaling applies to the typical value of the zero-temperature conductance $G_{RL}(V)$, see Eq.~\eqref{eq:exp_small} \footnote{{Evaluation of $G_{RL}(V)$ with the help of the finite-temperature version of Landauer formula and Eq.~\eqref{eq:exp_small} reveals a new bias scale for the behavior of the conductance, $V_\star = T L / (v\tau)$, associated with the temperature $T$.
At $V > V_\star$, a finite temperature produces negligible corrections to our results. At $V < V_\star$, on the other hand, $G_{RL}(V)$ is dominated by rare thermal fluctuations in which the charge is transmitted by carriers with energy close to the Fermi level, irrespective of the bias $V$.}}.}
It is useful to compare the typical value ${\cal T}_{\rm typ}(E)$ to an \textit{average} one, ${\cal T}_{\rm av} = 4 \langle e^{-2\lambda}\rangle$. Using Eq.~\eqref{eq:distr_props}, we find
\begin{equation}\label{eq:T_av}
    {\cal T}_{\rm av}(E) \sim \exp\Bigl[- \frac{3L}{v\tau \ln^2\bigl(\frac{1}{E\tau}\bigr)}\Bigr]
\end{equation}
(also see Ref.~\onlinecite{gogolin1998}).
Clearly, ${\cal T}_{\rm av}(E) \gg {\cal T}_{\rm typ}(E)$ in the considered low energy limit, $E \ll 1/\tau$. The reason is that---in contrast to the typical transmission---${\cal T}_{\rm av}(E)$ is determined by rare disorder configurations for which the transmission is nearly perfect, i.e., $\lambda \lesssim 1$. The difference between the average and the typical is standard for the Anderson localization problem in one dimension (see, e.g., Ref.~\onlinecite{abrikosov81}). The unusual aspect of the problem at hand is that the decay decrement of ${\cal T}_{\rm av}(E)$ differs from that of ${\cal T}_{\rm typ}(E)$ by a large, energy-dependent factor $\ln (1 / E \tau)$.

In the previous work \cite{kurilovich2022}, we found that the {\it local} conductance of a critical Majorana wire at bias $V = E / e$ is determined by its segment of size $l_{\rm cor}(E) =v \tau \ln^{2} (1 / E\tau)$.
Equation~\eqref{eq:T_av} elucidates the significance of this correlation length from the point of view of the transmission \textit{across} the wire. It shows that $l_{\rm cor}(E)$ determines the decay of the \textit{average} transmission coefficient with $L$.

\subsection{Localization away from the critical point\label{sec:disorder_away}}
Having discussed the localization properties at the critical point, we now turn to finding $l(E) \equiv l(E, B)$ at $B \neq B_{\rm c}$. 
To start with, we evaluate the low- and high-energy limits of $l(E, B)$.
At the Fermi level, $E = 0$, the particle-hole transformation demands that the backscattering phase vanishes, $\vartheta = 0$.
Application of Eq.~\eqref{eq:loc_length} then immediately yields:
\begin{equation}\label{eq:loc_Fermi}
l(0, B) = v / |\Delta_0|.
\end{equation}
We recall that $\Delta_0 \propto B - B_{\rm c}$, i.e., its value determines how close the system is to the topological transition point. Thus, $l(0, B) \propto |B - B_{\rm c}|^{-1}$ [see the inset of Fig.~\ref{fig:summary}(d)]. The critical exponent $-1$ here is {independent of disorder}. The divergence of $l(0, B)$ at the critical point is in agreement with the result of Sec.~\ref{sec:disorder_crit}.

Next, we find $l(E, B)$ in the opposite high-energy limit, $E \gg \tau^{-1}, |\Delta_0|$. Using the same approach as the one leading to Eq.~\eqref{eq:crit_pert}, we obtain 
\begin{equation}\label{eq:l_E_high_Delta_0}
    l(E, B) = 2 v \tau \Bigl(1 - \frac{\Delta_0^2 - 1/4\tau^2}{E^2} \Bigr).
\end{equation}
Regardless of the closeness to the transition point, $l(E, B)$ saturates at $l(E, B) = 2v\tau$ with $E$.

On the other hand, the way in which the value $l(E, B) = 2v\tau$ is approached does depend on the closeness to the critical point.
Equation~\eqref{eq:l_E_high_Delta_0} allows us to ``sharply'' define the notion of the {\it critical region} by the condition $|\Delta_0| \tau < 1 / 2$, see Fig.~\ref{fig:summary}(d). Within the critical region,  the localization length is a decreasing function of $|E|$. % {  (we recall that $E > 0$ in the present discussion)}.
The behavior is opposite outside of the critical region; there, $l(E, B)$ is the smallest at the Fermi level and increases monotonically with $|E|$. At the boundary of the critical region, $|\Delta_0| \tau = 1/2$, the localization length becomes independent of energy, see Appendix \ref{sec:special_point}.\\
%\vtext{Maybe $E$ instead of $|E|$?}

It is useful to note that the dependence of $l(E, B)$ on $E$, $\Delta_0 \propto B - B_{\rm c}$ and $1/\tau$ can be cast in the scaling form:
\begin{equation}\label{eq:scaling_law}
    l(E, B) = 2v\tau \cdot F(E\tau, \Delta_0 \tau).
\end{equation}
Using Eqs.~\eqref{eq:loc_Fermi} and \eqref{eq:l_E_high_Delta_0}, we obtain two limits of the scaling function $F(\varepsilon, \delta)$:
\begin{subequations}
\begin{align}
F(0, \delta) & = 1/(2|\delta|),\\
F(\varepsilon, \delta) & = 1 - [{\delta^2 - 1/4}]/{\varepsilon^2},\quad \varepsilon \gg |\delta|, 1.
\end{align}
\end{subequations}
In fact, it is possible to derive an integral representation for $F(\varepsilon, \delta)$ valid for arbitrary $\varepsilon$ and $\delta$.
To this end, we first solve Eq.~\eqref{eq:fokker-planck_st} for the steady-state distribution function of the backscattering phase $P(\vartheta)$, see Eqs.~\eqref{eq:distribution_exact} and \eqref{eq:prob_current} in Appendix~\ref{sec:general_l_E}. Then, we use the derived $P(\vartheta)$ to evaluate the harmonics in Eq.~\eqref{eq:loc_length}, see Eq.~\eqref{eq:l_E_exact}. 
We use the latter equation to plot $l(E, B)$ in Fig.~\ref{fig:summary}(e).

The boundary of the critical region separates the opposite trends in the dependence of $l(E, B)$ on $E$. Similarly, it separates the opposite trends in the behavior of the wire's density of states (DOS) $\nu(E)$. The monotonic decrease of $l$ with $E$ inside the critical region is accompanied by the DOS divergent at $E=0$,
\begin{equation}\label{eq:dos_low_energy}
    \nu(E) \propto \frac{1}{E^{1 - 2|\Delta_0| \tau}},
\end{equation}
see Ref.~\onlinecite{brouwer2011-2} and Appendix~\ref{sec:dos_derivation}. Outside the critical region, DOS develops a ``soft'' gap, cf.~Eq.~\eqref{eq:dos_low_energy} with $|\Delta_0|\tau >1/2$, concurrent with the change in the monotonicity of $l(E)$. 
At the boundary of the critical region, $|\Delta_0| \tau = 1 / 2$, the DOS $\nu(E)$ turns flat (see Appendix \ref{sec:dos_derivation}), similar to the behavior of $l(E)$. 

A large detuning from the critical point, $|\Delta_0|\tau \gg 1$, leads to the formation of a relatively well-pronounced ``gap'' of magnitude $E_{\rm gap} = |\Delta_0|$, see Fig.~\ref{fig:loc_away_intro}(b). The DOS becomes close to that of a clean wire $\nu_0(E)$,
\begin{equation}\label{eq:dos_0}
    \nu_0(E) = \frac{1}{\pi v} \frac{E}{\sqrt{E^2 - \Delta_0^2}}\Theta(E - |\Delta_0|).
\end{equation}
The only noticeable deviations occur near the gap edge: the ``coherence'' peaks are smeared by the disorder over an interval of width $\sim |\Delta_0| / (|\Delta_0|\tau)^{2 / 3} \ll |\Delta_0|$ \cite{ovchinnikov1977}. {In this large detuning regime, $|\Delta_0|\tau \gg 1$, it is possible to describe the crossover between asymptotes in Eqs.~\eqref{eq:loc_Fermi} and \eqref{eq:l_E_high_Delta_0} analytically. We do that in the next section.}

\subsection{Localization length for $|\Delta_0| \tau \gg 1$ \label{sec:disorder_strong}}
Here we find the energy dependence of the localization length for a relatively strong detuning of the field from the critical point, $|\Delta_0|\tau \gg 1$.

With respect to the localization length, there is no difference between trivial and topological phases. Therefore, we assume in the following that $\Delta_0 > 0$.

\subsubsection{${l(E)}$ at the above-the-gap energies}
We start by finding $l(E)$ at the above-the-gap energies, $E > \Delta_0$.
Were the wire clean, all states would be extended at such energies. 
Disorder, however, leads to their Anderson localization. 
Not too close to the ``band'' edge $E = \Delta_0$, the localization length $l(E)$ coincides with the perturbatively computed mean free path (up to a factor of $2$) \cite{thouless73}. This allows one to circumvent the general machinery of Sec.~\ref{sec:disorder} in evaluating $l(E)$ at $E > \Delta_0$.  To find the mean free path, we first  compute the disorder-averaged backscattering rate $\tau_{\rm sc}^{-1}(E)$ via Fermi Golden rule:
\begin{equation}\label{eq:rate}
    \frac{1}{\tau_{\rm sc}(E)} = 2\pi\,\frac{\nu_0(E)}{2} \frac{v}{\tau}.
\end{equation}
Here $\nu_0(E) / 2$ is the density of final states in the scattering process [Eq.~\eqref{eq:dos_0}].
The mean free path $l(E) / 2$ is related to the scattering rate via $v(E)\tau_{\rm sc}(E) = l(E) / 2$, where $v(E) = v \sqrt{E^2 - \Delta_0^2} / E$ is the velocity of a quasiparticle at energy $E$.
Combining this relation with Eq.~\eqref{eq:rate}, we find
\begin{equation}\label{eq:loc_away_above}
    l(E) = 2v\tau \bigl[1 - \Delta_0^2/E^2\bigr].
\end{equation}
We show in Appendix~\ref{sec:l_E_above_general} how this result can be obtained using the general formalism of Sec.~\ref{sec:disorder} [i.e., Eqs.~\eqref{eq:fokker-planck_st} and \eqref{eq:loc_length}].

{The localization length approaches $l(E) = 2v\tau$ at high energies [cf.~Eq.~\eqref{eq:l_E_high_Delta_0}], and decreases with the decreases of $E$.
Close to the gap edge, $E - \Delta_0 \ll \Delta_0$, $l(E)$ depends linearly on $E - \Delta_0$:
\begin{equation}\label{eq:l_E_close_to_gap}
l(E) = 4 v\tau (E - \Delta_0) / \Delta_0 \ll v\tau.
\end{equation}
%At lower energies, $0 < E - \Delta_0 \lesssim \Delta_0$,
At such energies, the presence of the gap strongly affects the density of states $\nu_0(E)$ and the quasiparticle velocity $v(E)$. Specifically, $\nu_0(E)$ increases as $E$ approaches $\Delta_0$ (the ``coherence'' peaks), whereas $v(E)$ becomes smaller. Both effects result in the decrease of $l(E)$ with $E - \Delta_0$.}

The perturbative calculation leading to Eqs.~\eqref{eq:loc_away_above}--\eqref{eq:l_E_close_to_gap} is valid provided the quasiparticle wavelength $2\pi / k(E)$ is short compared to the localization length $l(E)$, i.e., $k(E) l(E) \gg 1$.
%, where $k(E)$ is the quasiparticle wavevector. 
Sufficiently close to the gap edge, $E = \Delta_0$, this condition breaks down.
We can estimate the width of the energy interval in which the breakdown happens by setting $k(E) l(E) \sim 1$. 
The wavevector here can be approximated by $k(E) = \sqrt{2\Delta_0(E - \Delta_0)} / v$.
Combining the latter expression with Eq.~\eqref{eq:l_E_close_to_gap},
we see that Eq.~\eqref{eq:loc_away_above} becomes inapplicable at
\begin{equation}\label{eq:E_star}
    E - \Delta_0 \sim E_{\star},\quad E_{\star} =  \Delta_0 / (\sqrt{2}\Delta_0 \tau)^{2/3}
\end{equation}
(for convenience, we introduced the factor of $\sqrt{2}$ in the definition of $E_{\star}$). 
Note that $E_\star \ll \Delta_0$ in the considered regime of $\Delta_0 \tau \gg 1$. In fact, $E_{\star}$ coincides with the energy scale that determines the smearing of the coherence peaks in the density of states \cite{ovchinnikov1977}. We now obtain the localization length $l(E)$ for energies close to the gap edge, $|E - \Delta_0| \lesssim E_\star$.

\subsubsection{$l(E)$ near and below the gap edge}
Because the perturbation theory fails near the gap edge, $|E - \Delta_0| \lesssim E_\star$, one has to resort to general equations \eqref{eq:fokker-planck_st} and \eqref{eq:loc_length} to find $l(E)$. We obtain $l(E)$ in two steps. First, we evaluate the distribution function $P(\vartheta)$ of the backscattering phase. Then, we use $P(\vartheta)$ to perform the averaging in Eq.~\eqref{eq:loc_length}.

\begin{figure}[t]
  \begin{center}
    \includegraphics[scale = 1]{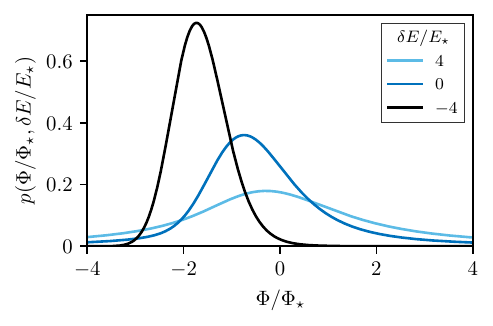}
    \caption{Distribution function of the backscattering phase for different values of $\delta E \equiv E - \Delta_0$ [the curves are produced using Eq.~\eqref{eq:distr_dimless}]. Variable $\Phi$ used in the plot is related to backscattering phase $\vartheta$ via $\Phi \equiv \vartheta + \pi / 2$. Parameter $E_\star$ is defined in Eq.~\eqref{eq:E_star}; $\Phi_\star$ is defined after Eq.~\eqref{eq:Phi_star}.}
    \label{fig:distribution}
  \end{center}
\end{figure}

Let us begin by finding $P(\vartheta)$. As we described above, one can visualize the evolution of $\vartheta$ with $x$ as that of a particle sliding down the washboard potential $U_{\rm eff}(\vartheta) = - E \vartheta + \Delta_0 (\cos \vartheta - 1)$. For energies $|E - \Delta_0| \ll \Delta_0$, the sliding particle spends the majority of ``time'' close to the point $\vartheta = -\pi / 2$, where the drift velocity $\propto (E + \Delta_0 \sin \vartheta)$ is nearly vanishing [cf.~Eq.~\eqref{eq:backscattering}]. 
In other words, the phase distribution function $P(\vartheta)$ is concentrated in a narrow interval $|\vartheta + \pi/2| \ll 1$.
This suggests parameterizing $\vartheta = -\pi / 2 + \Phi$ and performing an expansion in $\Phi \ll 1$ in Fokker-Planck equation~\eqref{eq:fokker-planck_st}. We obtain the following equation for the steady state distribution function:
\begin{equation}\label{eq:P_steady_half_pie}
    \frac{1}{\Delta_0 \tau} \frac{\partial^2 P}{\partial \Phi^2} - \frac{1}{2}\frac{\partial}{\partial \Phi}\Bigl(\Bigl[\Phi^2 + \frac{2\delta E}{\Delta_0}\Bigr]P\Bigr) = 0,
\end{equation}
where $\delta E \equiv E - \Delta_0$.
One can bring Eq.~\eqref{eq:P_steady_half_pie} to a convenient form depending on a single parameter by performing a rescaling:
\begin{equation}\label{eq:Phi_star}
    \Phi = \Phi_{\star}\cdot\varphi,\quad \quad\delta E = E_{\rm \star}\cdot \varepsilon.
\end{equation}
Here, $\Phi_{\star} = (2/\Delta_0\tau)^{1/3}$ and $E_\star$ is defined in Eq.~\eqref{eq:E_star}.
Note that $\Phi_\star \ll 1$ and $E_\star \ll \Delta_0$ in the considered limit of $\Delta_0 \tau \gg 1$.
In terms of the rescaled variables, Eq.~\eqref{eq:P_steady_half_pie} acquires the form 
\begin{equation}
    \frac{\partial^2 P}{\partial {\varphi}^2} - \frac{\partial}{\partial {\varphi}} \bigl[ {\varphi}^2 + \varepsilon\bigr]P = 0.
\end{equation}
Its solution that decays at $|\varphi| \gg 1$ and has a constant probability current reads
\begin{equation}\label{eq:distr_MIR}
    P(\Phi, E) = \Phi_{\star}^{-1} p(\varphi, \varepsilon) \equiv \Phi_{\star}^{-1} p\Bigl(\frac{\Phi}{\Phi_{\star}}, \frac{\delta E}{E_{\star}}\Bigr),
\end{equation}
where $p(\varphi, \varepsilon)$ is a normalized by unity function, $\int d\varphi\,p(\varphi, \varepsilon) = 1$, given by 
\begin{equation}\label{eq:distr_dimless}
    p({\varphi}, \varepsilon) = {\cal N} \int_{{\varphi}}^{+\infty} e^{\frac{{\varphi}^3}{3} - \frac{y^3}{3} + \varepsilon ({\varphi} - y)}dy,
\end{equation}
where ${\cal N}$ is the normalization constant.
According to Eq.~\eqref{eq:distr_MIR}, the distribution function is confined to an interval $|\Phi|\lesssim \Phi_{\star}$ for energies $|\delta E| \lesssim E_{\star}$. Function $p(\Phi / \Phi_\star, \delta E / E_\star)$ is depicted in Fig.~\ref{fig:distribution}.

We now use distribution \eqref{eq:distr_MIR} to evaluate $l(E)$. We will verify shortly that---for the considered energies $|E - \Delta_0| \ll \Delta_0$---the second term in Eq.~\eqref{eq:loc_length} is negligible in comparison with the first one. Therefore, we can approximate
${l^{-1}(E)} = -\frac{\Delta_0}{v} \langle \cos \vartheta \rangle \equiv -\frac{\Delta_0}{v} \langle \sin \Phi \rangle$,
where we used $\vartheta \equiv \pi/2 + \Phi$. In addition, it follows from Eqs.~\eqref{eq:distr_MIR} and \eqref{eq:distr_dimless} that $\Phi \ll 1$ for $|E - \Delta_0| \ll \Delta_0$. We can thus expand the sine, which yields
\begin{equation}\label{eq:loc_MIR}
    \frac{1}{l(E)} = -\frac{\Delta_0}{v} \langle \Phi\rangle.
\end{equation}
Using Eq.~\eqref{eq:distr_MIR} in Eq.~\eqref{eq:loc_MIR},  we arrive at
\begin{align}\label{eq:scaling}
    l(E) = v\tau &\cdot \frac{E_{\star}}{\Delta_0}\cdot f\Bigl(\frac{\delta E}{E_{\star}}\Bigr),\notag\\
    &f(\varepsilon) = \Bigr[\int d{\varphi}(-{\varphi}) \,p({\varphi}, \delta E / E_{\star})\Bigr]^{-1}
\end{align}
with $p({\varphi}, \varepsilon)$ of Eq.~\eqref{eq:distr_dimless}.
By evaluating the integral in Eq.~\eqref{eq:scaling}, we obtain an explicit form of the scaling function $f(\varepsilon)$:
\begin{equation}\label{eq:scaling_2}
    f(\varepsilon) = 2 \Bigl(-\frac{\partial}{\partial \varepsilon} \ln \bigl[{\rm Ai}^2(-\varepsilon) + {\rm Bi}^2(-\varepsilon)\bigr]\Bigr)^{-1},
\end{equation}
where ${\rm Ai}(x)$ and ${\rm Bi}(x)$ are the two kinds of Airy functions. Function $f(\varepsilon)$ is depicted in the inset of Fig.~\ref{fig:loc_away_intro}(a).

We note in passing that, in the considered energy interval $|E - \Delta_0| \ll \Delta_0$, the quasiparticle energy dispersion can be approximated by a quadratic one, $E(k) = \Delta_0 + v^2 k ^2 / (2\Delta_0)$.
Therefore, expectedly, the obtained results for the localization length [Eqs.~\eqref{eq:scaling}, \eqref{eq:scaling_2}] coincide with the respective results for a $p^2/2m$-particle (of mass $m = \Delta_0 / v^2$) moving in the short-range-correlated Gaussian random potential \cite{lifshitz1988}.

Let us now analyze Eq.~\eqref{eq:scaling}. First, we consider the case $E = \Delta_0$, i.e., $\delta E = 0$. We obtain
\begin{align}\label{eq:at_the_edge_exact}
    l(E) = \frac{\Gamma (1/6)}{3^{1/3}\sqrt{\pi}}\,\frac{v\tau}{(\Delta_0 \tau)^{2/3}} \approx 2.2\,v\tau / (\Delta_0 \tau)^{2/3},
\end{align}
where we used Eq.~\eqref{eq:E_star} to express $E_\star$ in terms of $\Delta_0$ and $\tau$.
We see that $l(E)\ll v\tau$ in the considered limit of $\Delta_0\tau \gg 1$. This justifies the made omission of the second term in Eq.~\eqref{eq:loc_length}, which is $\sim 1/v\tau \ll 1/l(E)$. 

While the main purpose of Eq.~\eqref{eq:scaling} is to describe the behavior of $l(E)$ in the non-perturbative energy interval $|E - \Delta_0| \lesssim E_{\star}$, it is, in fact, applicable in a wider range, $|E - \Delta_0| \ll \Delta_0$. 
It is useful to apply Eq.~\eqref{eq:scaling} in the intermediate region $E_{\star} \ll |\delta E| \ll \Delta_0$. Using the asymptotes of Airy functions together with Eq.~\eqref{eq:scaling}, we find
\begin{subequations}
    \begin{align}\label{eq:asym_1}
       l(E) &= l_0\cdot\frac{4\delta E}{\Delta_0},\quad\quad \delta E \gg E_{\star},\\ 
       l(E) &= \frac{v}{\sqrt{2\Delta_0|\delta E|}},\quad |\delta E| \gg E_{\star},\quad  \delta E < 0. \label{eq:asym_2}
    \end{align}
\end{subequations}
Equation \eqref{eq:asym_1} agrees with the derived earlier Eq.~\eqref{eq:l_E_close_to_gap}. 

We note that the localization length becomes independent of disorder at ``subgap'' energies, see Eq.~\eqref{eq:asym_2}.
The found $l(E)$ agrees with the expansion of the respective clean-limit expression, see Eq.~\eqref{eq:xi_ind}.
Using the latter equation, we can generalize Eq.~\eqref{eq:asym_2} to all subgap energies as
\begin{equation}\label{eq:l_subgap}
    l(E) = \frac{v}{\sqrt{\Delta_0^2 - E^2}}.
\end{equation}
We discuss in Appendix~\ref{sec:l_E_subgap_general} how this expression can be obtained directly from Eqs.~\eqref{eq:fokker-planck_st} and \eqref{eq:loc_length}.

The behavior of $l(E)$ for the wire strongly detuned from the critical point, $\Delta_0 \tau \gg 1$, is summarized in Fig.~\ref{fig:loc_away_intro}(a).

\section{Microscopic model of a Majorana wire\label{sec:micro}}
Here, we exemplify the presented universal theory with a concrete microscopic model of a Majorana wire. To this end, we consider a by-now standard model of a proximitized single-band quantum wire with Rashba spin-orbit coupling \cite{lutchyn2010, oreg2010}. Using it as a starting point{---and focusing on a limit of strong spin-orbit coupling---}we show how the low-energy theory of Secs.~\ref{sec:low-energy_theory}--\ref{sec:disorder} can be derived, and find its parameters $v, \Delta_0$, and $\tau$. 
We also show how deviations from the predicted universal behavior arise in the case of strong disorder and for energies comparable to the proximity-induced pairing potential. % $\Delta_{\rm ind}$. 
{  
While simple, the model of this section makes a number of assumptions inapplicable to experiments \cite{aghaee2023}. 
We present the results of a numerical simulation for realistic material parameters in the next section [Sec.~\ref{sec:for_real}].}

We assume that a single-band quantum wire is proximitized by an $s$-wave superconductor and subjected to a parallel magnetic field. The system can be described by the following many-body Hamiltonian:
\begin{equation} \label{eq:mbody}
    H_{\rm wire} = \frac{1}{2}\int dx \, \Psi^\dagger(x) \hat{H} \mathrm \Psi(x),
\end{equation}
where $\Psi = \bigl(\psi_\uparrow,\,\psi_\downarrow,\,\psi^\dagger_\downarrow,\,-\psi^\dagger_\uparrow \bigr)^T$, $\psi_\sigma$ is the annihilation operator of electrons with spin $\sigma = \, \uparrow$ or $\downarrow$, and
\begin{equation}\label{eq:NW}
    \hat{H} = \Bigl(\frac{p^2}{2m} - \mu + \alpha p \sigma_z + V(x)\Bigr) \tau_z + \Delta_{\rm ind} \tau_x  - B \sigma_x.
\end{equation}
Here, $\sigma_{i}$ and $\tau_{i}$ with $i\in \{x, y, z\}$ are the Pauli matrices in spin space and Nambu space, respectively, $p = -i\partial_x$, $\mu$ is the chemical potenital, $m$ is the effective mass, $\alpha$ is the spin-orbit coupling constant, $\Delta_{\rm ind}$ is the proximity-induced pairing potential, and $B$ is the Zeeman energy. We neglect the orbital effect of the magnetic field. 
Lastly, $V(x)$ is the disorder potential. We assume that $V(x)$ is a Gaussian random variable with a zero average, $\langle V(x) \rangle = 0$, and a short-ranged correlation function,
\begin{equation}\label{eq:corr_function_micro}
    \langle V(x) V(x^\prime) \rangle = \frac{\alpha}{\tau_{\rm ns}}\,\delta(x - x^\prime).
\end{equation}
We expressed the correlation function in terms of the (Drude) mean free time in the normal state $\tau_{\rm ns}$. 

To simplify the problem, we focus on the regime of strong spin-orbit coupling, $m\alpha^2 \gg \mu, B, \Delta$. In it, the spectrum of the wire near the Fermi level consists of two pairs of helical modes that are well-separated in the momentum space. One pair resides close to momenta $p \approx 0$ (``inner'' modes) while the other pair resides near $p \approx \pm 2m\alpha$ (``outer'' modes). The momentum separation allows one to split the electron field operator into its helical components:
\begin{equation}\label{eq:decomp_helical}
    \psi_\sigma(x) = \psi_{{\rm i}, \sigma}(x) + \psi_{{\rm o}, \sigma}(x) e^{-2 i \sigma m \alpha x},
\end{equation}
where $\psi_{\rm i, \sigma}(x)$ and $\psi_{\rm o, \sigma}(x)$ are the annihilation operators for an electron in the inner and outer mode, respectively. 
In contrast to $\psi_\sigma(x)$---which oscillates with a wavelength $\lambda_F \sim 1/m\alpha$---fields $\psi_{{\rm i},\sigma}(x)$ and $\psi_{{\rm o},\sigma}(x)$ change slowly with $x$; the scale of their variation is $\gtrsim \alpha / \Delta_{\rm ind} \gg \lambda_F$~\footnote{Hereafter, we assume that parameters $\mu$ and $B$ are of the same order as $\Delta_{\rm ind}$.}. Substituting the decomposition \eqref{eq:decomp_helical} into Eq.~\eqref{eq:NW} and coarse graining the Hamiltonian over distances $\gtrsim \lambda_F$, we arrive to
\begin{align}\label{eq:NW_cg}
    &H_{\rm wire} = \frac{1}{2}\sum_{s = {\rm i},{\rm o}}\int dx\, \Psi^\dagger_{s}(x) \hat{H}_s \Psi_s(x)\notag\\
    &+\int dx\, \Bigl(\Psi^\dagger_{\rm i}(x) \frac{V_1(x) - i\sigma_z V_2(x)}{2\sqrt{2}}\tau_z \Psi_{\rm o}(x) + {\rm h.c.}\Bigr),
\end{align}
where $\Psi_{\rm s} = \bigl(\psi_{\rm s, \uparrow}, \psi_{\rm s, \downarrow}, \psi^\dagger_{\rm s, \downarrow}, -\psi^\dagger_{\rm s, \uparrow}\bigr)^T$ and
\begin{subequations}\label{eq:H_io}
\begin{align}
    \hat{H}_{\rm i} &= [\alpha p \sigma_z - \mu + V_0(x)]\tau_z - B \sigma_x + \Delta_{\rm ind}  \tau_x,\label{eq:H_i}\\
    \hat{H}_{\rm o} &= [-\alpha p \sigma_z - \mu + V_0(x)]\tau_z + \Delta_{\rm ind}  \tau_x.\label{eq:H_o}
\end{align}
\end{subequations}
Note that in the considered limit of large spin-orbit coupling, $m \alpha^2 \gg B$, only the inner helical modes are affected by the magnetic field.
Fields $V_{0}(x)$ and $V_{1,2}(x)$ in Eqs.~\eqref{eq:NW_cg} and \eqref{eq:H_io} describe forward scattering and backscattering by the disorder potential, respectively. They are mutually independent Gaussian random variables satisfying
\begin{equation}
    \langle V_m(x) V_n(x^\prime)\rangle = \frac{\alpha}{\tau_{\rm ns}}\,\delta_{mn} \delta(x - x^\prime),
\end{equation}
where $m, n \in \{0, 1, 2\}$.

\begin{figure*}[t]
  \begin{center}
    \includegraphics[scale = 1]{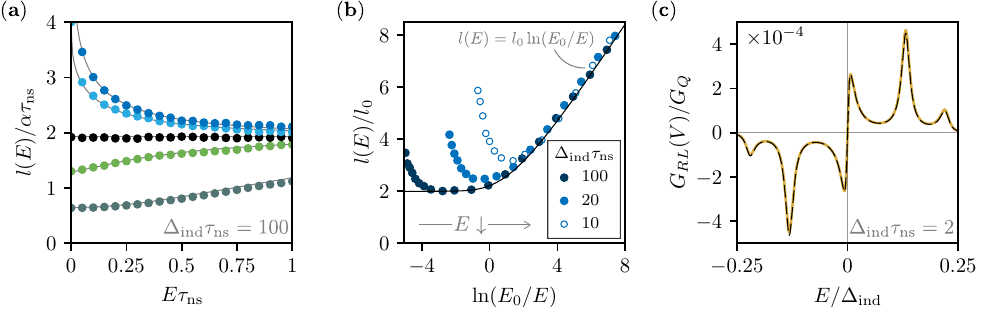}
    \caption{
    {\bf a.} Energy dependence of the localization length $l(E)$ near the critical point. The numerically found $l(E)$ is in agreement with the analytically evaluated universal behavior (grey lines) at low energies and weak disorder,
    %for weak disorder, 
    $\tau_{\rm ns}^{-1} = 0.01\Delta_{\rm ind}$. 
    Data points are produced by finding the average Lyapunov exponent $\langle \lambda(E , L)\rangle$ %numerically 
    for a wire of length $L = 100\cdot \alpha \tau_{\rm ns}$ with $\mu = \Delta_{\rm ind}$.
    Dark blue dots correspond to the critical point, $B=B_c$. For the chosen parameters, the critical region width is $\delta B = 0.38\,\tau_{\rm ns}^{-1}$. Other data sets correspond to $B - B_{\rm c} = \delta B\cdot \{0.5, 1, 1.5, 3\}$.
    {\bf b.} The $l(E)$ dependence at the critical field in a broader range of energies, $E\lesssim\Delta_{\rm ind}$, and for a range of disorder strengths. Deviations from the universal low-energy theory (solid line) are evident at higher energies.
    {\bf c.} Example of $G_{RL}(V)$ for one realization of strong disorder, $\tau_{\rm ns}^{-1} = 0.5 \Delta_{\rm ind}$ [$L = 10\cdot \alpha \tau_{\rm ns}$, $\mu = \Delta_{\rm ind}$; field $B=B_{\rm c}$]. Dashed line depicts $-G_{RL}(-V)$ and illustrates the symmetry of the $G_{RL}(V)$ function. 
    }
    \label{fig:disorder_numerics}
  \end{center}
\end{figure*}

In the absence of disorder, the outer and the inner helical modes decouple. The topological transition occurs at \cite{lutchyn2010, oreg2010}
\begin{equation}\label{eq:Bc_clean}
B_\mathrm{c} = (\Delta^2_{\rm ind} + \mu^2)^{1/2},
\end{equation}
and is associated with the closing of the gap in the spectrum of the inner modes. 
The low-energy degrees of freedom near the gap closing are a pair of Majorana fermions $\chi_R(x)$ and $\chi_L(x)$ [cf.~Fig.~\ref{fig:summary}(b)]. An explicit relation between Majorana fields $\chi_{\alpha}(x)$ and Dirac fields $\psi_{{\rm i}, \sigma}(x)$ can be established by diagonalizing the single-particle Hamiltonian in Eq.~\eqref{eq:H_i}. We find
\begin{subequations}\label{eq:Majoranas}
\begin{align}
    \hspace{-0.2cm}\chi_R(x) &= u_\mu \frac{\psi_{{\rm i}, \uparrow}(x)\hspace{-0.05cm} - \hspace{-0.05cm}\psi^\dagger_{{\rm i}, \uparrow}(x)}{\sqrt{2}i} - v_\mu \frac{\psi_{{\rm i},\downarrow}(x) \hspace{-0.05cm}-\hspace{-0.05cm} \psi^\dagger_{{\rm i},\downarrow}(x)}{\sqrt{2}i},\\
    \hspace{-0.2cm}\chi_L(x) &= u_\mu \frac{\psi_{{\rm i}, \downarrow}(x) \hspace{-0.05cm}+\hspace{-0.05cm} \psi^\dagger_{{\rm i},\downarrow}(x)}{\sqrt{2}} - v_\mu \frac{\psi_{{\rm i}, \uparrow}(x) \hspace{-0.05cm}+ \hspace{-0.05cm}\psi^\dagger_{{\rm i}, \uparrow}(x)}{\sqrt{2}}.
\end{align}
\end{subequations}
Here the $uv$-factors read
\begin{subequations}
\begin{align}
    u_\mu &= \frac{1}{\sqrt{2}}\Bigl[1 + \frac{\Delta_{\rm ind} }{\sqrt{\Delta^2_{\rm ind}  + \mu^2}}\Bigr]^{1/2},\\
    v_\mu &= \frac{{\rm sgn}\,\mu}{\sqrt{2}}\Bigl[1 - \frac{\Delta_{\rm ind} }{\sqrt{\Delta^2_{\rm ind}  + \mu^2}}\Bigr]^{1/2}.
\end{align}
\end{subequations}
At $B = B_{\rm c}$, the many-body Hamiltonian of a clean wire can be represented in terms of $\chi_{\alpha}(x)$ as
\begin{equation}\label{eq:NW_crit}
    H_{\rm wire, c} = -\frac{i}{2}\int dx\,v(\chi_R \partial_x \chi_R - \chi_L \partial_x \chi_L) + \dots,
\end{equation}
where $\dots$ encodes contributions involving the high-energy modes (i.e., the modes with $E \gtrsim \Delta_{\rm ind}$). The velocity is the same for the left- and right-moving Majorana fermions, and is given by
\begin{equation}\label{eq:velocity_result}
    v = \alpha \frac{\Delta_{\rm ind} }{\sqrt{\Delta^2_{\rm ind}  + \mu^2}},
\end{equation}
as follows directly from Eq.~\eqref{eq:H_i}.

Next, we account for the detuning from the critical point and (weak) disorder. To this end, we project the full Hamiltonian \eqref{eq:NW_cg} onto the low-energy subspace [cf.~Eq.~\eqref{eq:Majoranas}]. A straightforward calculation yields the effective Hamiltonian
\begin{equation}\label{eq:H_eff}
    H_{\rm eff} = H_{\rm wire, c} + \delta H_{\rm wire} + H_{\rm dis}.
\end{equation}
Here $H_{\rm wire, c}$ is given by Eq.~\eqref{eq:NW_crit}. $\delta H_{\rm wire}$ describes the detuning from the critical point and is given by Eq.~\eqref{eq:detuning} with
\begin{equation}\label{eq:Delta_0}
    \Delta_0 = B - B_{\rm c}.
\end{equation}
Term $H_{\rm dis}$ in Eq.~\eqref{eq:H_eff} describes the disorder. To the first order in $V_m(x)$, it is given by Eq.~\eqref{eq:disorder} with 
\begin{equation}\label{eq:deltaDelta_micro}
\delta \Delta(x) = V_0(x) \frac{\mu}{\sqrt{\Delta^2_{\rm ind} + \mu^2}}.
\end{equation}
Notice that $\delta \Delta(x)$---which gives the \textit{backscattering} amplitude of the Majorana fermions---is, in fact, determined by the \textit{forward} scattering component of the microscopic potential $V(x)$. 
Combining Eqs.~\eqref{eq:corr_function_micro}, \eqref{eq:velocity_result}, and \eqref{eq:deltaDelta_micro}, we can bring the correlation function of $\delta \Delta(x)$ to the form of Eq.~\eqref{eq:correlator}, with the mean free time
\begin{equation}\label{eq:tau_micro}
    \frac{1}{\tau} = \frac{1}{\tau_{\rm ns}}\,\frac{\mu^2}{\Delta_{\rm ind} B_{\rm c}}.
\end{equation}
Interestingly, the right hand side vanishes at $\mu = 0$. This is the feature of the lowest-order calculation in the disorder strength. At $\mu = 0$, the scattering of the Majorana fermions is governed by the processes of the second order in $V_m(x)$  \cite{brouwer2011-2}.

The above perturbative treatment is justified provided the disorder is relatively weak, $\Delta_{\rm ind} \tau_{\rm ns} \gg 1$. 
To the leading order in a small parameter $(\Delta_{\rm ind} \tau_{\rm ns})^{-1}$, the critical field coincides with its clean-limit value, see Eq.~\eqref{eq:Bc_clean}. In fact, though, disorder leads to a renormalization of $B_{\rm c}$ in the next-to-the-leading order  by $\delta B_{\rm c} / B_{\rm c} \sim  (\Delta_{\rm ind}\tau_{\rm ns})^{-1}$~\cite{brouwer2011-2}.

For arbitrary disorder strength and away from the low-energy limit, one can evaluate the transport properties of the wire numerically. We find the scattering matrix for the Hamiltonian~\eqref{eq:NW_cg} following the approach of Ref.~\onlinecite{brouwer2011-2}. The scattering matrix allows us to determine the conductance $G_{RL}(V)$ and extract the localization length $l(E)$. 

As an illustration, we first use the numerical simulation to reaffirm the predictions of the low-energy theory for $l(E)$ in the case of a %moderately 
weak disorder and $E \ll \Delta_{\rm ind}$, see Fig.~\ref{fig:disorder_numerics}(a). 
The simulations also allow us to find $l(E)$ for a range of disorder strengths and to quantify the deviations from the low-energy theory at $E \sim \Delta_{\rm ind}$, see Fig.~\ref{fig:disorder_numerics}(b, c).

At $B = B_{\rm c}$, the localization length diverges logarithmically at $E \rightarrow 0$ [see~Eq.~\eqref{eq:l_E_crit}]. This singular behavior is a universal feature of the infinite randomness criticality, and is valid regardless of the ``microscopic'' disorder strength, see Fig.~\ref{fig:disorder_numerics}(b). 
The low-energy theory predicts that the singularity is followed by the monotonic decrease of $l(E)$ with $E$. 
Figure~\ref{fig:disorder_numerics}(b) shows that this prediction breaks down at $E \sim \Delta_{\rm ind}$, where the monotonicity of $l(E)$ changes [also see Sec.~\ref{sec:for_real}].

Finally, let us recall another central prediction of the low-energy theory: close to $B = B_{\rm c}$, $G_{RL}(V)$ is odd in $V$ at low biases [see Eq.~\eqref{eq:G_RL_P} and related discussion]. We can use the numerical simulation to assess how well this prediction holds in the regime of the strong disorder. Figure~\ref{fig:disorder_numerics}(c) demonstrates that $G_{RL}(V) = -G_{RL}(-V)$ remains valid even for $\Delta_{\rm ind} \tau_{\rm ns} \sim 1$, as long as $|V| \ll \Delta_{\rm ind} / e$.

\section{Results of a numerical simulation for realistic material parameters\label{sec:for_real}}
{  What remains to show is that our conclusions about the behavior of the localization length $l(E)$ and the antisymmetry of the nonlocal conductance are applicable for realistic material parameters.
In this Section, we present numerical results for %the behavior of 
$l(E)$
and $G_{RL}(V)$ near the topological transition for values of spin-orbit coupling and disorder potential relevant to experimentally studied InAs nanowires \cite{aghaee2023}.

Specifically, we focus on the implementation of a Majorana wire based on a gate-defined single-subband InAs quantum wire proximitized by a superconducting aluminum layer. 
We describe the quantum wire by the Hamiltonian \eqref{eq:NW}, in which we take $m = 0.03 m_{\rm e}$ for the effective electron mass in InAs ($m_{\rm e}$ is the bare electron mass), $\alpha = 10\,{\rm meV}\cdot {\rm nm}$ for the Rashba spin-orbit coupling, and $\Delta_{\rm ind} = 150\,\mu{\rm eV}$ for the (zero-field) induced pairing potential \cite{aghaee2023}.
We note that for such parameters the energy scale associated with the spin-orbit coupling is small, $m \alpha^2 \lesssim \Delta_{\rm ind}$. This is in contrast to a strong assumption $m \alpha^2 \gg \Delta_{\rm ind}$ made in Sec.~\ref{sec:micro}.
The chemical potential $\mu$ is a control parameter set by the voltage applied to the back gate of the device; we take $\mu = 500\,\mu{\rm eV}$. 
Finally, we assume that the disorder potential $V(x)$ is a Gaussian random variable with the correlation function $\langle V(x) V(x^\prime)\rangle = V_{\rm rms}^2 \exp(- |x - x^\prime| / r_{\rm c})$. We take 
$V_{\rm rms} = {500}\,\mu {\rm eV}$, on par with the disorder potential in 
the cleanest wires available \cite{aghaee2023}. The correlation radius, taken to be $r_{\rm c} = 100\,{\rm nm}$, is associated with the characteristic distance between charge impurities in the heterostructure, and the electron gas.
}

\subsection{Localization length $l(E)$}

\begin{figure*}[t]
  \begin{center}
    \includegraphics[scale = 1]{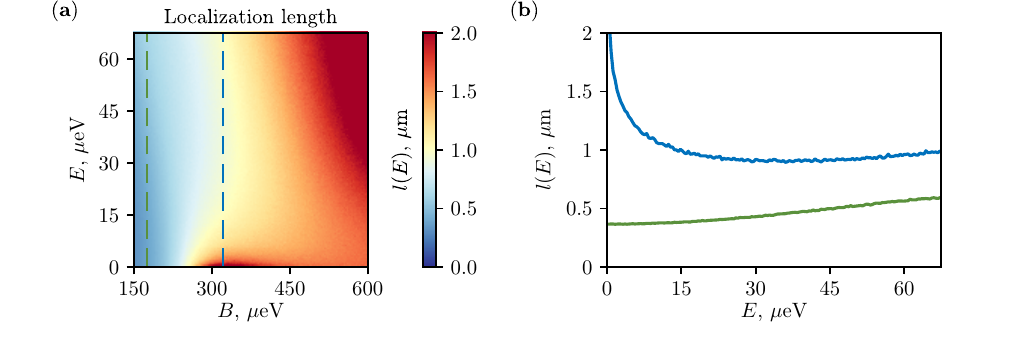}
    \caption{{\bf a.} Evolution of the energy-dependence of the localization length $l(E)$ with the variation of the Zeeman energy $B$. The critical point of the topological transition is $B_{\rm c} \approx 321\,\mu{\rm eV}$. {\bf b.} $l(E)$ for two values of $B$ depicted in panel {\bf a} with vertical dashed lines. Blue curve corresponds to the critical point, $B = B_{\rm c}$.
    The green curve corresponds to the value $B \approx 175\,\mu {\rm eV}$ outside of the critical region in the trivial phase.}
    \label{fig:numerical_data}
  \end{center}
\end{figure*}
{  {To numerically obtain the localization length $l(E)$ here, we discretize Hamiltonian~\eqref{eq:NW} by writing derivatives as finite differences (lattice parameter $a=1$ nm ensures the absence of discretization artifacts), implement the solution of the Bogoliubov-de Gennes equation by means of transfer matrix, and follow a standard procedure (see, e.g., Ref.~\onlinecite{mackinnon1983}) for calculating the transfer-matrix eigenvalue spectrum. Next, we determine the 
asymptotic length scaling of the eigenvalues. This scaling defines the Lyapunov exponents and, thus, the localization length $l(E)$.}
The evolution of $l(E)$ with the Zeeman energy $B$ is shown in Fig.~\ref{fig:numerical_data}. 
It confirms the validity of the universal low-energy theory conclusions for the realistic wire parameters, in which the spin-orbit coupling is weak and disorder is strong.
At the critical point, the localization length has a logarithmic singularity at the Fermi level, $l(E) \propto \ln (1 / E)$ [blue curve in Fig.~\ref{fig:numerical_data}(b)]. Tuning $B$ sufficiently strongly away from the critical point---i.e., outside of the critical region---changes the monotonicity of $l(E)$ at small $E$ {[green curve in Fig.~\ref{fig:numerical_data}(b)]}. 
}

\subsection{Comparison of antisymmetric and symmetric parts of $G_{RL}(V)$}
\begin{figure*}[t]
  \begin{center}
    \includegraphics[scale = 1]{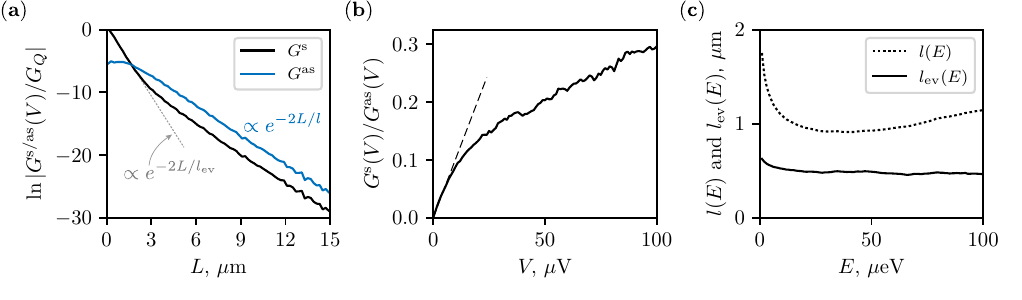}
    \caption{  
    \textbf{a:} Comparison of symmetric and antisymmetric components of the conductance at different wire lengths for realistic material parameters. We fix $V = 5\,{\rm \mu V}$ and find $\ln |G_{RL}^{\rm s/as}(V)|$ using Kwant simulation package \cite{groth_2014}; we average the result over ${100}$ disorder realizations. Zeeman energy is $B = B_{\rm c} \approx 321\,{\rm \mu eV}$. 
    {For a long wire, $L \gtrsim 3\,{\rm \mu m}$, the antisymmetric component exceeds the symmetric one; both components vary with length exponentially, with the same characteristic length $l$.
    At smaller $L$, the antisymmetric component saturates whereas the symmetric component crosses over to a different exponential dependence $\propto \exp(-2L/ l_{\rm ev})$, with $l_{\rm ev} < l$}.
    \textbf{b:} Dependence of $G^{\rm s}(V) / G^{\rm as}(V)$ on bias $V$ at $L = 12.5\,{\rm \mu m}$. 
    The dependence is linear at $V \rightarrow 0$ [dashed line shows a respective asymptote].
    \textbf{c:} Localization length $l(E)$ and the decay scale of evanescent modes $l_{\rm ev}(E)$. To extract $l(E)$ and $l_{\rm ev}(E)$, we fit {$\ln|G^{\rm s} / G_Q|$ to $\ln[g_1 \exp(-2L / l_1) + g_2\exp(-2L/l_2)]$, and identify $l_{\rm ev}(E)$ and $l(E)$ with ${\rm min}(l_1,l_2)$ and ${\rm max}(l_1,l_2)$, respectively.}
    }
    \label{fig:numerical_data_2}
  \end{center}
\end{figure*}

{  The low energy theory of Sec.~\ref{sec:disorder} predicts that the nonlocal conductance is an antisymmetric function of bias at $V\rightarrow 0$ for a sufficiently long wire. 
In realistic conditions, the antisymmetry can be violated by the energy dependence of the junctions transmission coefficients.
Additionally, electron transport via high-energy evanescent modes (relevant in the short-wire limit \cite{stanescu_2013}) also disrupts the conductance antisymmetry \cite{Danon2020}. 
To analyze and separate these two effects for realistic wire parameters, we perform numerical simulation using Kwant package \cite{groth_2014} for a set of bias voltages and wire lengths.

To start with, we split $G_{RL}(V)$ into symmetric (``s'') and antisymmetric (``as'') components $G_{RL}^{\rm s / as}(V) = [G_{RL}(V) \pm G_{RL}(-V)] / 2$.
%In order to capture the effect of evanescent modes,
We compare the dependence of $\ln (|G_{RL}^{\rm s}(V)| / G_Q)$ and $\ln (|G_{RL}^{\rm as}(V)| / G_Q)$ on the wire length at a fixed small bias of $V = 10\,\mathrm{\mu V}$, see Fig.~\ref{fig:numerical_data_2}(a).
For a long wire ($L \gtrsim 1.5\,{\rm \mu m}$), 
the antisymmetric and symmetric components decay exponentially with the same localization length, $l = 1.1\,{\rm \mu m}$; the asymmetric component exceeds the symmetric one.
We associate the extracted value of $l$ with the decay of the Majorana mode which dominates the conductance.

The fact that the Majorana mode contributes to the symmetric component (in addition to its major contribution to the antisymmetric one) is a manifestation of a finite bias effect stemming from the energy dependence of the junction transmission, see Eq.~\eqref{eq:G_RL_P}. 
Indeed, unlike the first and the third terms, the second term in Eq.~\eqref{eq:G_RL_P} does not have a well-defined parity; therefore,
\begin{equation}
    \frac{G^{\rm s}_{RL}(V)}{G^{\rm as}_{RL}(V)} = \frac{T_{\rm me}^L(E) - T_{\rm me}^L(-E)}{T_{\rm me}^L(E) + T_{\rm me}^L(-E)}\Bigl|_{E = eV}.
\end{equation}
Thus, we expect ${G^{\rm s}_{RL}(V)} / {G^{\rm as}_{RL}(V)} \propto V$ at small biases.
In Fig.~\ref{fig:numerical_data_2}(b), we show the numerically extracted dependence of ${G^{\rm s}_{RL}(V)} / {G^{\rm as}_{RL}(V)}$ on $V$ at fixed length $L = 12.5\,{\rm \mu m}$.
In agreement with the expectation, the dependence is linear at $V \rightarrow 0$.

Our next observation is the difference in the behavior of $G^{\rm s}_{RL}$ and $G^{\rm as}_{RL}$ at short wire lengths. The variation with $L$ of the asymmetric part saturates at $L\lesssim l$. 
In contrast, rather than saturating, the symmetric part crosses over to an exponential dependence with a distinct characteristic length, $l_{\rm ev} = 0.3\,{\rm \mu m}$ {[the respective asymptote is shown in Fig.~\ref{fig:numerical_data_2}(a) by a dotted line]}.
We associate this shorter characteristic length with the contribution of the evanescent modes, most prominent in the symmetric part of the conductance.
To further verify this, in Fig.~\ref{fig:numerical_data_2}(c), we show the dependence of $l(E)$ and $l_{\rm ev}(E)$ on $E$.
{We extract $l(E)$ and $l_{\rm ev}(E)$ by fitting  $\ln|G^{\rm s} / G_Q|$ to $\ln[g_1 \exp(-2L / l_1) + g_2\exp(-2L/l_2)]$, and identifying $l_{\rm ev}(E) = {\rm min}(l_1,l_2)$ and $l(E) = {\rm max}(l_1,l_2)$.}
As expected from the contribution of critical modes, $l(E)$ has a divergence at $E \rightarrow 0$. 
On the other hand, $l_{\rm ev}(E)$ is featureless in this limit; this is consistent with the contribution of non-critical, high-energy modes.}

\section{Discussion and conclusions}
We found the nonlocal conductance $G_{RL}(V)$ of a proximitized quantum wire near the critical point of the topological phase transition.

The low-energy degrees of freedom in the wire are a pair of counter-propagating Majorana fermions [see Eqs.~\eqref{eq:wire_c}--\eqref{eq:disorder}].
The neutral character of Majorana fermions makes $G_{RL}(V)$ an odd function at low biases $V$ [see Eqs.~\eqref{eq:G_RL_P} and \eqref{eq:G_RL_low_bias}]. 
The observation that the nonlocal conductance becomes an odd function of $V$ upon application of the magnetic field $B$ to the wire is qualitatively consistent with the recent observations \cite{puglia2021, banerjee2023}.

If the length of the wire $L$ is short compared to the (normal-state) mean free path, then one can disregard the influence of disorder. At the critical point, the dependence of $G_{RL}$ on $V$ of such a clean wire has a series of equidistant peaks 
%at $V = \pi v n / e L$ ($n \in \mathbbm{Z}$), where $v$ is the velocity of the Majorana fermions 
[see Eq.~\eqref{eq:clean_crit_PE} and Fig.~\ref{fig:cond_clean}(a)].  The peaks originate from the Fabry-P\'erot interference.
Detuning of $B$ from its critical value $B_{\rm c}$ opens up a ``transport'' gap in the bias dependence of the conductance, see Fig.~\ref{fig:cond_clean}(b). Below the gap, $G_{RL}(V)$ is suppressed exponentially in $L$ [see Eq.~\eqref{eq:FM}].
For the clean wire, 
the gap extracted from the transport measurement coincides with the spectral gap $E_{\rm gap} \propto |B - B_{\rm c}|$. 

The disorder cannot be neglected for a long wire. 
Due to the disorder-induced backscattering, 
%In contrast to the clean case,
the conductance of a ``dirty'' wire becomes exponentially small in $L$ at {\it all} biases $V$ (as opposed to $G_{RL}(V)$ of a clean wire).
Decay of the typical value of the conductance with $L$ occurs over the scale of the quasiparticle localization length $l(E = eV)$, which we found [see Sec.~\ref{sec:disorder}].

The qualitative character of $l(E)$ changes as the wire is tuned across $B = B_{\rm c}$.
Away from the critical point, $l(E)$ is an increasing function of energy ($E$ is measured with respect to the Fermi level), see Figs.~\ref{fig:summary}(d) and \ref{fig:loc_away_intro}(a) and Eqs.~\eqref{eq:loc_away_above}, \eqref{eq:scaling}, \eqref{eq:scaling_2}, and \eqref{eq:l_subgap}.
Such behavior of $l(E)$ reflects gradual weakening of the localization effect with the increase of electron energy. On the contrary, the trend is opposite 
%This changes 
in a narrow vicinity of $B_{\rm c}$:
%, where the trend is opposite; 
the quasiparticle becomes {\it less} localized at low-energies, see Figs.~\ref{fig:summary}(d) and Eq.~\eqref{eq:l_E_high_Delta_0}.
Exactly at $B = B_{\rm c}$, the low-energy behavior of the localization length is singular, $l(E) \propto \ln(1/E)$, see Eq.~\eqref{eq:l_E_crit}. 
This reflects the nature of the criticality, which belongs to the infinite-randomness universality class \cite{motrunich2001, fisher1994}. 
We find a good agreement between the analytical low-energy theory and numerical simulations for the Rashba wire model (see Sec.~\ref{sec:micro} and \ref{sec:for_real}). {  The agreement holds for the 
%This includes the case of 
experimentally relevant values \cite{aghaee2023} of disorder potential and spin-orbit coupling, provided the wire is sufficiently long (see Sec.~\ref{sec:for_real}).
The requirement on the length is set by the decay scale of the 
%decay length-scale 
of evanescent, high-energy modes in the wire.}

The behavior of $l(E)$ is the same on both sides of the phase transition. Nevertheless, nonlocal measurements allow one to distinguish a topological wire from a trivial wire. We predict that the presence of Majorana zero modes in the topological phase leads to low-bias resonances in the transport characteristics of the wire. The electroneutrality of Majorana modes makes these resonances exponentially small in $L$ in the case of the conductance measurements, see Eq.~\eqref{eq:resonant}. However, the resonances are much more pronounced in the non-local noise, see Eqs.~\eqref{eq:noise_resonant}--\eqref{eq:Fano}.

{While the results for the noise in Sec.~\ref{sec:majoranas} are obtained in the clean limit, they remain applicable in the presence of weak disorder.
In the latter case, the Majorana splitting $\varepsilon_M$ in Eqs.~\eqref{eq:noise_resonant}--\eqref{eq:Fano} becomes a random quantity determined by a disorder realization in a given sample.
The statistical properties of $\varepsilon_M$ are known \cite{brouwer2011}. Away from the critical region, the distribution of $\varepsilon_M$ is log-normal with $\langle \ln (\varepsilon_M / \Delta_0) \rangle = - L \Delta_0 / v$ and ${\rm var}\ln (\varepsilon_M / \Delta_0) = L / v \tau$.} 

Our results for the conductance are directly applicable to nonlocal transport measurements such as those reported in Refs.~\onlinecite{puglia2021, banerjee2023, aghaee2023}.
The localization length $l(E)$ can be accessed in a device with a series of contacts along the wire allowing one to vary the effective length~$L$ [see, e.g., Fig.~27 of Ref.~\onlinecite{aghaee2023}].
The qualitative difference in the energy dependence of $l(E)$ close to and away from the critical point can be used as a signature of the topological phase transition. 

\acknowledgements{
This research was supported at Yale University by the Office of Naval Research (ONR) under award number N00014-22-1-2764, by the Army Research Office (ARO) under grant number W911NF-22-1-0053, and by the NSF Grant No. DMR-2410182.}

\appendix

\begin{figure*}[t]
  \begin{center}
    \includegraphics[scale = 1]{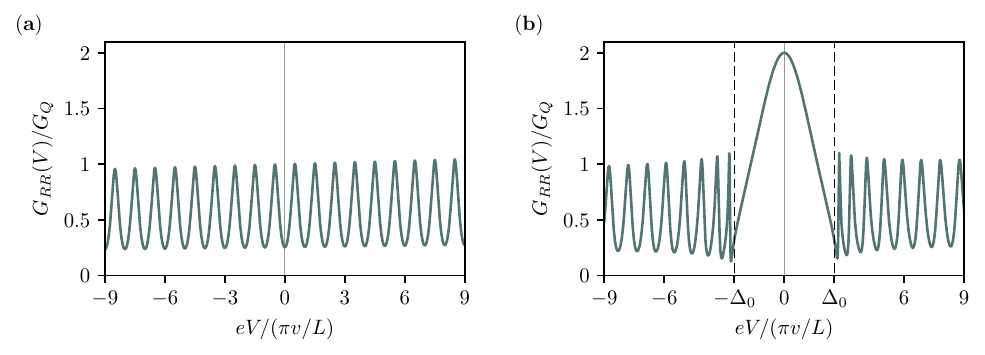}
    \caption{Local differential conductance $G_{RR}(V)$ of a clean wire at (panels {\bf a}) and away from (panel {\bf b}) the critical point at zero temperature. We use the same parameters as in Fig.~\ref{fig:cond_clean}.
    Presence of the Majorana zero mode at the wire's end results in a peak
    in the local conductance of a quantized height, $G_{RR}(0) = 2G_Q$.}
    \label{fig:local}
  \end{center}
\end{figure*}

\section{Derivation of $S^M(E)$ in Eq.~\eqref{eq:Sm_solution}\label{sec:SM_derivation}}
In this Appendix, we present a derivation of the scattering matrix of Majorana modes $S^M(E)$ across a clean wire tuned away from the critical point, see Eq.~\eqref{eq:Sm_solution}. Finding the $S$-matrix amounts to solving the Schr\"odinger equation defined by Eqs.~\eqref{eq:wire_c} and \eqref{eq:detuning}. 
Assuming that the wire spans an interval $x \in [0, L]$, whereas $x < 0$ and $x > L$ are ``wave regions'' [see the discussion in Sec.~\ref{sec:general} and Fig.~\ref{fig:waves}], we can represent the single-particle wavefunction at energy $E$ as 
\begin{align}
    &\psi(x)=\\
    &\begin{cases}
        m_{\rm r} 
        \begin{pmatrix}
            1\\0
        \end{pmatrix} e^{ik x} + m_{\rm l} 
        \begin{pmatrix}
            0\\1
        \end{pmatrix} e^{-ik x},\quad x < 0,\\
        q_{\rm d}
        \begin{pmatrix}
            {\rm sgn}\,\Delta_0\\-i e^{i\eta}
        \end{pmatrix} e^{-\frac{L}{l}} + q_{\rm g} 
        \begin{pmatrix}
            {\rm sgn}\,\Delta_0\\-i e^{-i\eta}
        \end{pmatrix} e^{\frac{L}{l}},\quad 0 < x < L,\\
        \tilde{m}_{\rm r} 
        \begin{pmatrix}
            1\\0
        \end{pmatrix} e^{ik x} + \tilde{m}_{\rm l} 
        \begin{pmatrix}
            0\\1
        \end{pmatrix} e^{-ik x},\quad x > L.
    \end{cases}\notag
\end{align}
Here $k(E) = E / v$, $l(E)$ is given by Eq.~\eqref{eq:xi_ind}, and $\eta(E)$ is defined in Eq.~\eqref{eq:eta}.
By demanding the continuity of the wavefunction at $x = 0$ and $x = L$, and solving the resulting system of equations for $\tilde{m}_{\rm r}$, $m_{\rm l}$, $q_{\rm d}$, and $q_{\rm g}$, we arrive to Eq.~\eqref{eq:Sm_solution} for $S^M(E)$.

\section{``Gauge'' fixing of the Majorana modes\label{sec:gauge}}
Even though the Majorana fermions are described by real fields (as opposed to complex fields describing Dirac fermions), there is still a $\mathbbm{Z}_2$ ``gauge'' freedom under the transformation $\chi_{i}\rightarrow -\chi_{i}$. Throughout the manuscript, we fix the gauge in a particular way which we describe here.

The particle-hole symmetry [cf.~Eq.~\eqref{eq:ph_junctions}] demands that the reflection amplitude of a Majorana fermion off a right junction is real at the Fermi level, $S^R_{\rm mm}(0)\equiv r_{\rm m}^R(0) = [r_{\rm m}^R(0)]^\star$. The sign of the amplitude is not fixed and depends on the gauge choice. We choose the gauge in such a way that the reflection off a closed junction happens with an amplitude $r^R_{\rm m}(0) = + 1$. 

In fact, this choice uniquely determines that the phase with $\Delta_0 > 0$ is the topological one. Indeed, the topological phase is characterized by the presence of a Majorana zero mode at the wire's ends. The Bohr-Sommerfeld quantization condition for the presence of a $E = 0$ bound state at the right end reads $i\,{\rm sgn}\,\Delta_0 \cdot e^{i\eta(0)}\cdot r^R_{\rm m}(0) = +1$ [as follows from Eqs.~\eqref{eq:Sr_def} and \eqref{eq:Sm_solution}]. We see that for $r^R_{\rm m}(0) = +1$ it is satisfied for $\Delta_0 > 0$.

In addition to a Majorana zero mode at the right end, a closed topological wire ($\Delta_0 > 0$) hosts a Majorana zero mode at the left end. This imposes a consistency constraint on the reflection amplitude off the left junction $S^L_{\rm mm}(0) \equiv - r^L_{\rm m}(0)$ [note the minus sign]. The Bohr-Sommerfeld condition for the presence of a Majorana zero mode at the left end reads  $-i\,{\rm sgn}\,\Delta_0 \cdot e^{i\eta(0)}\cdot (-r^L_{\rm m}(0)) = +1$. This fixes $r^L_{\rm m}(0) = +1$ for a closed junction.

\section{Local conductance\label{sec:local}}
The approach developed in Sec.~\ref{sec:general} can also be used to find the {\it local} differential conductance $G_{RR}(V)$ measured at the end of the wire. 
At zero temperature, the latter quantity can be expressed as \begin{equation}G_{RR}(V) = G_Q ( |r^R_{\rm ee}(E)|^2 - |r^R_{\rm he}(E)|^2 )\bigr|_{E = eV},
\end{equation}
where $r^R_{\rm ee}(E)$ and $r^R_{\rm he}(E)$ are reflection amplitudes for an electron impinging on the wire from the right lead in normal and Andreev channels, respectively. The reflection amplitudes can be obtained by solving the scattering problem of Fig.~\ref{fig:waves} with $a_{\rm e / h} = 0$, $d_{\rm e} = 1$, and $d_{\rm h}$; in this case, $r^R_{\rm ee / he} \equiv c_{\rm e / h}$. An example of $G_{RR}(V)$ resulting from the solution of the scattering problem is shown in Fig.~\ref{fig:local}.

\section{Derivation of Eq.~\eqref{eq:diff_noise} for the differential noise\label{sec:diff_noise_derivation}}
In this Appendix, we derive Eq.~\eqref{eq:diff_noise} for the differential noise $\partial {\cal S}_{RL}(V) \partial V$. In general, the noise can be expressed in terms of the transmission amplitudes in particle and hole channels in the following way (see, e.g., Ref.~\onlinecite{akhmerov2011}):
\begin{equation}\label{eq:noise}
\frac{\partial {\cal S}_{RL}(V)}{\partial V} = \frac{e^3}{2\pi}\bigl[{\cal T}_+(eV) - {\cal T}_-^2(eV)],
\end{equation}
where 
\begin{equation}
    {\cal T}_\pm(E) = |S_{\rm ee}(E)|^2 \pm |S_{\rm he}(E)|^2.
\end{equation}
Substituting Eq.~\eqref{eq:she} into the latter equation, we obtain \begin{equation}
{\cal T}_{\pm}(E) = \bigl(T^R_{{\rm em}}(E) \pm T^R_{{\rm hm}}(E)\bigr){T_{{\rm me}}^L(E) Q(E)}.
\end{equation}
%Here $Q(E)$ is the same function as in Eq.~\eqref{eq:G_RL_P}; it describes propagation of a Majorana mode from the left junction to the right.
For low biases, we can further expand the transmission coefficients of the junctions to the lowest non-vanishing order in $E$. This leads to
\begin{subequations}
\begin{align}
    {\cal T}_+(E) &= 2T_{\rm em}^R(0) T_{\rm me}^L(0) Q(E),\label{eq:T_+}\\
    {\cal T}_-(E) &= \frac{E}{\Sigma_R}\cdot T_{\rm em}^R(0) T_{\rm me}^L(0) Q(E),
\end{align}
\end{subequations}
where $\Sigma_R$ is defined in Eq.~\eqref{eq:Sigma_R}.
Clearly, ${\cal T}_-(E) \ll {\cal T}_+(E)$ at energies $E \ll \Sigma_R$. Thus, at such energies, one can neglect the second term in the square brackets in Eq.~\eqref{eq:noise}. Then, using Eq.~\eqref{eq:T_+},  we arrive to Eq.~\eqref{eq:diff_noise}. 

\section{Derivation of system \eqref{eq:dynamics_finite_E} for $\vartheta$ and $\lambda$\label{sec:derivation_evol}} 
In this Appendix, we present details of the derivation of system \eqref{eq:dynamics_finite_E} for the evolution of backscattering phase $\vartheta$ and Lyapunov exponent $\lambda$.

As explained in Sec.~\ref{sec:disorder}, to find the scattering matrix of Majorana modes, we break the wire into a sequence of short elements (each of which can be treated perturbatively), and then track how the $S$-matrix evolves as the elements are joined together. The $S$-matrix relates the incoming and outgoing waves according to
\begin{equation}\label{eq:Sm_def_app}
\begin{pmatrix}
\tilde{m}_{\rm r}\\m_{\rm l}
\end{pmatrix}
=
{\widetilde{S}}^M(E,x)
\begin{pmatrix}
m_{\rm r} \\ \tilde{m}_{\rm l}
\end{pmatrix}.
\end{equation}
Note that this definition differs from the one used in the main text by the presence of phase factors [cf.~Eqs.~\eqref{eq:Sm_def_app} and \eqref{eq:Sm_def}]. Such a definition is more convenient for the derivation; we highlight the difference by a tilde sign.

The scattering matrix of a short element of length $\delta x$ can be found with the help of the perturbation theory. Using Born approximation, we obtain
\begin{align}\label{eq:Sm_add_dx}
    {\widetilde{S}^M_{\delta x}(E, x)} &
    = 1 - i \begin{pmatrix}
    0 & {\delta {\cal U}^\star(x)}\\
    {\delta {\cal U}(x)} & 0
    \end{pmatrix},
\end{align}
where
\begin{equation}
    \delta {\cal U}(x) = -{i}e^{2ik(E) x}
    \int_x^{x+\delta x}dx^\prime \frac{\Delta(x^\prime)}{v},
\end{equation}
wavevector $k(E) = E / v$, and $\Delta(x) = \Delta_0 + \delta \Delta (x)$.

An addition of the short element to the wire of length $x$ changes the $S$-matrix of the latter by $\delta \widetilde{S}^M(E, x)$, i.e., $\widetilde{S}^M(E, x + \delta x) = \widetilde{S}^M(E, x) + \delta \widetilde{S}^M(E, x)$. Using Eq.~\eqref{eq:Sm_add_dx}, we find for $\delta \widetilde{S}^M(E, x)$ to the first order in $\delta {\cal U}(x)$:
\begin{align}\label{eq:s_evol_finite_diff}
\begin{split}
    i\delta {\widetilde{S}}_{\rm rr}^M &=  \delta {\cal U}\, {\widetilde{S}}^M_{\rm rl} {\widetilde{S}}^M_{\rm rr},\\
    i\delta {\widetilde{S}}_{\rm ll}^M &=  \delta {\cal U}\, {\widetilde{S}}^M_{\rm rl} {\widetilde{S}}^M_{\rm ll},\\
    i\delta {\widetilde{S}}^M_{\rm rl} &=  \delta {\cal U}^\star + \delta {\cal U}\, [{\widetilde{S}}^M_{\rm rl}]^2,\\
    i\delta {\widetilde{S}}^M_{\rm lr} &=  \delta {\cal U}\, {\widetilde{S}}^M_{\rm ll} {\widetilde{S}}^M_{\rm rr}.
\end{split}
\end{align}
Here ${\widetilde{S}}^M_{ij}$ with $i,j \in \{{\rm r}, {\rm l}\}$ denotes the components of the $S$-matrix.

Finally, we perform two additional transformations. First, we promote finite-difference equations \eqref{eq:s_evol_finite_diff} to differential equations. Second, we relate $\widetilde{S}^M(E,x)$ defined via Eq.~\eqref{eq:Sm_def_app} to ${S}^M(E,x)$ [cf.~Eq.~\eqref{eq:Sm_def}]. This amounts to setting $\widetilde{S}^M_{\rm rr / ll} (E, x) = e^{-ik(E)x}{S}^M_{\rm rr / ll} (E, x)$, $\widetilde{S}^M_{\rm rl} (E, x) = e^{-2ik(E)x}{S}^M_{\rm rl} (E, x)$, and $\widetilde{S}^M_{\rm lr} (E, x) = {S}^M_{\rm lr} (E, x)$. As a result, we obtain
\begin{align}
\begin{split}
    \frac{d{S}_{\rm rr}^M}{dx} &= i\frac{E}{v}{S}_{\rm rr}^M -  \frac{\Delta(x)}{v} \, {{S}}^M_{\rm rl} {{S}}^M_{\rm rr},\\
    \frac{d{S}_{\rm ll}^M}{dx} &=
    i\frac{E}{v}{S}_{\rm ll}^M -  \frac{\Delta(x)}{v}\, {{S}}^M_{\rm rl} {{S}}^M_{\rm ll},\\
    \frac{d{S}_{\rm rl}^M}{dx} &=
    i\frac{2E}{v}{S}_{\rm rl}^M -  \frac{\Delta(x)}{v} \bigl( [{{S}}^M_{\rm rl}]^2 - 1\bigr),\\
    \frac{d{S}_{\rm lr}^M}{dx} &=  -  \frac{\Delta(x)}{v}\, {{S}}^M_{\rm ll} {{S}}^M_{\rm rr}.
\end{split}
\end{align}
Using parameterization \eqref{eq:S_param} in these equations, we arrive to Eq.~\eqref{eq:dynamics_finite_E}.

\section{Derivation of the Lyapunov exponent distribution at the critical point\label{sec:lyap_disr_appendix}}
Here, we consider a wire tuned to the critical point and find the distribution function $\Pi_0(\lambda)$ of the Lyapunov exponent $\lambda$. Throughout the Appendix, we focus on the low-energy regime, $E \ll 1/ \tau$.

To start with, we use Eqs.~\eqref{eq:correlator} and \eqref{eq:dynamics_finite_E} to derive a Fokker-Planck equation for the evolution of the joint distribution function $P$ of Lyapunov exponent $\lambda$ and backscattering phase $\vartheta$. In doing this, we use a convenient $w$-parameterization for the backscattering phase, $w = \ln \tan (\vartheta / 2)$. Following a standard approach (see, e.g., Ref.~\onlinecite{yamamoto2021}), we find for $P(\lambda, w|x)$:
\begin{equation}\label{eq:FP_total}
v\tau\frac{\partial P}{\partial x}
= \frac{1}{2} \tanh^2w \frac{\partial^2 P}{\partial \lambda^2} + \hat{L}[w, E] P + \hat{V}[w]\frac{\partial P}{\partial \lambda}.
\end{equation}
Here $\hat{L}[w, E]$ is a linear operator governing the diffusion of the $w$-variable. Its action on a function $f(w)$ is defined by 
\begin{equation}
    \hat{L}[w, E]f(w) = 2\frac{\partial^2 f(w)}{\partial w^2} - 2 E \tau \frac{\partial}{\partial w}\bigl(\cosh w f(w)\bigr).
\end{equation}
Linear operator $V[w]$ in Eq.~\eqref{eq:FP_total} acts according to
\begin{equation}\label{eq:V_operator}
    \hat{V}[w] f(w) = \tanh w \frac{\partial f(w)}{\partial w} + \frac{\partial}{\partial w}\bigl(\tanh w f(w)\bigr).
\end{equation}
It describes the coupling between $\lambda$ and $w$. we now explain how one can ``integrate out'' the dynamics of variable $w$ and, in this way, obtain an equation for the evolution of distribution function $\Pi_0(\lambda|x) = \int dw\,P(\lambda, w|x)$ of the Lyapunov exponent.

To set the stage for evaluation of $\Pi_0(\lambda|x)$, we first establish the spectral structure of $\hat{L}[w,E]$. Note that the form of this operator coincides with the diffusion kernel for a Brownian particle moving in an external potential $U_{\rm eff}(w) =-2E\tau \sinh w$. The structure of the potential is remarkably simple in the considered low-energy limit, $E \ll 1 / \tau$, see the discussion between Eqs.~\eqref{eq:FP_w} and \eqref{eq:distribution_crit_low}. 
The dynamics of the $w$-``particle'' primarily occurs within an equipotential interval $w \in [-\ln (1 / E\tau), \ln (1/ E\tau)]$. There, the action of $\hat{L}[w,E]$ reduces to
\begin{equation}\label{eq:simple_diffusion}
    \hat{L}[w, E] = 2\frac{\partial^2}{\partial w^2}.
\end{equation}
We can account for the structure of $U_{\rm eff}(w)$ outside of the equipotential region by imposing appropriate boundary conditions on the space of functions $f(w)$ within which operator $\hat{L}[w,E]$ acts. For $w \gtrsim \ln(1/E\tau)$, $U_{\rm eff}(w)$ rapidly drops without a bound. We can account for this by demanding
\begin{equation}\label{eq:cliff_bc}
    f\bigl(\ln(1/E\tau)\bigr) = 0.
\end{equation}
When the particle falls down the ``cliff'', $w \rightarrow +\infty$, it reemerges on the opposite side of the $w$-line, as required by the continuity of the $S$-matrix evolution. After that, the external field quickly (in ``time'' $\Delta x \sim v \tau$) sweeps it to a point $w = -\ln (1 / E\tau)$. The conservation of the probability current requires
\begin{align}\label{eq:current_cons_bc}
    \frac{\partial f}{\partial w}\Bigr|_{w = \ln(\frac{1}{E\tau})} = \frac{\partial f }{\partial w}\Bigr|_{w = -\ln(\frac{1}{E\tau})}
\end{align}
Equation \eqref{eq:simple_diffusion} defined within an interval $w \in [-\ln(1/E\tau), \ln(1/E\tau)]$ and supplemented by boundary conditions \eqref{eq:cliff_bc} and \eqref{eq:current_cons_bc} contains full information on spectral properties of $\hat{L}[w,E]$ at energies $E \ll 1/\tau$.

Let us now find the eigenvalues and the respective eigenfunctions of $\hat{L}[E,w]$. Zero mode $f_0(w)$ of $\hat{L}[E,w]$ describes the steady state of the $w$-particle diffusion. We find the zero mode by demanding $\hat{L}[w,E]f_0(w) = 0$. This leads to
\begin{equation}\label{eq:zero_mode}
    f_0(w) = \frac{1}{2\ln(\frac{1}{E\tau})}\Bigl(1 - \frac{w}{\ln\bigl(\frac{1}{E\tau}\bigr)} \Bigr),
\end{equation}
see~Eq.~\eqref{eq:distribution_crit_low}. One can enumerate all other eigenfunctions by a positive integer $n$. These functions satisfy $\hat{L}[w,E]f_n(w) = \Lambda_n f_n(w)$ with $\Lambda_n \neq 0$. From this equation, we obtain
\begin{equation}\label{eq:f_n_Lambda_n}
    f_n(w) = \sqrt{\frac{2}{\ln(\frac{1}{E\tau})}} \sin\Bigl[\frac{\pi n w}{\ln(\frac{1}{E\tau})}\Bigr],\quad\Lambda_n = -2\Bigl[\frac{\pi n}{\ln(\frac{1}{E\tau})}\Bigr]^2
\end{equation}
(the choice of the normalization coefficient is arbitrary at this point; the only requirement for the normalization is condition \eqref{eq:normalization} presented below).

An unusual aspect of the considered problem is that operator $\hat{L}[w, E]$ is, in fact, {\it non-diagonalizable}; it has Jordan blocks of size two. A function $\tilde{f}_n(w)$ belonging to the $n$-th Jordan chain satisfies
\begin{equation}\label{eq:Jordan}
    {\hat{L}}[w, E]\tilde{f}_n(w) = \Lambda_n \tilde{f}_n(w) + \widetilde{\Lambda}_n f_n(w),
\end{equation}
where $f_n(w)$ and $\Lambda_n$ are given by Eq.~\eqref{eq:f_n_Lambda_n}.
By solving Eq.~\eqref{eq:Jordan}, we find
\begin{equation}
    \tilde{f}_n(w) = \frac{1 - \frac{w}{\ln(\frac{1}{E\tau})}}{\sqrt{2\ln(\frac{1}{E\tau})}}\cos\Bigl[\frac{\pi n w}{\ln(\frac{1}{E\tau})}\Bigr]
\end{equation}
and $\widetilde{\Lambda}_n = \pi n / \ln^2\bigl(\frac{1}{E\tau}\bigr)$.

In addition to $f_n(w)$ and $\tilde{f}_n(w)$, we introduce a set of adjoint functions $\tilde{f}_n^t(w)$ and $f_n^t(w)$. These functions are, respectively, the eigenfunctions and the Jordan chain functions of the transposed operator $\hat{L}^T[w, E]$. By solving $\hat{L}^T[w, E] \tilde{f}_n^t(w) = \Lambda_n \tilde{f}_n^t(w)$ and $\hat{L}^T[w, E] f_n^t(w) = \Lambda_n f_n^t(w) + \widetilde{\Lambda}_n \tilde{f}_n^t(w)$, we obtain for $n \geq 1$:
\begin{subequations}
\begin{align}
f^t_n(w) &= \frac{1 + \frac{w}{\ln(\frac{1}{E\tau})}}{\sqrt{2 \ln(\frac{1}{E\tau})}} \sin \Bigl[\frac{\pi n w}{\ln(\frac{1}{E\tau})}\Bigr],\\
\tilde{f}_n^t(w) &= \sqrt{\frac{2}{\ln(\frac{1}{E\tau})}} \cos \Bigl[\frac{\pi n w}{\ln(\frac{1}{E\tau})}].
\end{align}
\end{subequations}
The zero mode of $\hat{L}^T[w, E]$ is given by \begin{equation}\label{eq:zero_mode_adj}
f^t_0(w) = 1.
\end{equation}
Adjoint functions $\tilde{f}_n^t(w)$ and $f_n^t(w)$ are orthonormal to $f_n(w)$ and $\tilde{f}_n(w)$:
\begin{subequations}\label{eq:normalization}
\begin{align}
(f_n^t|f_{n^\prime}) = (\tilde{f}_n^t|\tilde{f}_{n^\prime}) &= \delta_{nn^\prime},\\ (f_n^t|\tilde{f}_{n^\prime}) = (\tilde{f}_n^t|{f}_{n^\prime}) &= 0.
\end{align}
\end{subequations}
Here we defined the inner product of functions $\psi_1(w)$ and $\psi_2(w)$ according to
\begin{equation}
    (\psi_1|\psi_2) = \int dw\,\psi_1(w)\psi_2(w),
\end{equation}
where the integration is carried over an interval $w \in [-\ln(1/E\tau), \ln(1/E\tau)]$. This finishes the preliminaries needed to find the distribution function $\Pi_0(\lambda)$.

A set of functions $\{f_0(w), f_n(w), \tilde{f}_n(w)\}$ with $n\geq 1$ forms a basis. It is convenient to perform an expansion of the joint distribution function $P(\lambda, w|x)$ in this basis:
\begin{align}\label{eq:decomposition}
    P(\lambda,\,& w|x) = \Pi_0(\lambda|x) f_0(w)\notag \\ &+ \sum_{n = 1}^{\infty}\bigl[\Pi_n(\lambda|x)f_n(w) + \widetilde{\Pi}_n(\lambda|x) \tilde{f}_n(w)\bigr].
\end{align}
Here $\Pi_0(\lambda|x) = (f_0^t|P) = \int dw\,P(\lambda, w|x)$ is the distribution function of the Lyapunov exponent. {$\Pi_0(\lambda|x)$ is coupled to  harmonics $\Pi_n$ and $\widetilde{\Pi}_n$ with $n\geq 1$ by an operator $\hat{V}[w]$, see Eq.~\eqref{eq:V_operator} and discussion after it.}
Substituting decomposition \eqref{eq:decomposition} in Eq.~\eqref{eq:FP_total} and acting on both side of the equation with $(f_0^t|$, we obtain for $\Pi_0(\lambda|x)$:
\begin{align}\label{eq:Pi0_to_Pi_n}
    v\tau& \frac{\partial \Pi_0}{\partial x} = \frac{1}{2}\frac{\partial^2 \Pi_0}{\partial \lambda^2} + (f_0^t|\hat{V}|f_0) \frac{\partial \Pi_0}{\partial \lambda}\notag\\
    &+\sum_{n = 1}^{\infty} \Bigl\{(f_0^t|\hat{V}|f_n) \frac{\partial \Pi_n}{\partial \lambda} + (f_0^t|\hat{V}|\tilde{f}_n) \frac{\partial \widetilde{\Pi}_n}{\partial \lambda}\Bigr\}.
\end{align}
In arriving to this result, we used the fact that one can approximate $\tanh^2 w \approx 1$ in the low-energy limit, $E \ll 1/\tau$. 

{As it stands, equation \eqref{eq:Pi0_to_Pi_n} is not closed with respect to $\Pi_0$.
%due to its coupling to harmonics with $n \geq 1$.
However, one can obtain an approximate closed equation in the large ``time'' limit, $x \gg v \tau \ln^2 (1/E \tau)$.}
%, one can simplify Eq.~\eqref{eq:Pi0_to_Pi_n} further and obtain an approximate closed equation on $\Pi_0$. 
Indeed, because $f_0(w)$ is a zero mode of $\hat{L}[w, E]$, the large-time dynamics of $\Pi_0$ occurs in a much slower way than the dynamics of $\Pi_n$ and $\widetilde{\Pi}_n$. In fact, $\Pi_n$ and $\widetilde{\Pi}_n$ follow $\Pi_0$ adiabatically for $x \gg v\tau \ln^2 (1 / E\tau)$. We find using Eq.~\eqref{eq:FP_total}:
\begin{equation}
    \begin{pmatrix}
        \Pi_n\\
        \widetilde{\Pi}_n
    \end{pmatrix}
    = -\frac{1}{\Lambda_n^2}
    \begin{pmatrix}
        \Lambda_n & -\widetilde{\Lambda}_n\\
        0 & \Lambda_n
    \end{pmatrix}
    \begin{pmatrix}
        (f_n^t|\hat{V}|f_0)\\
        (\tilde{f}_n^t|\hat{V}|f_0)
    \end{pmatrix}
    \frac{\partial \Pi_0}{\partial \lambda}.
\end{equation}
Substituting these expressions in Eq.~\eqref{eq:FP_total}, we obtain
\begin{equation}\label{eq:lyap_distr_equation}
v\tau \frac{\partial \Pi_0}{\partial x} = {\cal D} \frac{\partial^2 \Pi_0}{\partial \lambda^2} - {\cal V}\frac{\partial \Pi_0}{\partial \lambda}.    
\end{equation}
This equation has a form of a usual diffusion equation with a drift term. The drift velocity is given by 
\begin{equation}\label{eq:V_final}
    {\cal V} = -(f_0^t|\hat{V}|f_0) = \frac{1}{\ln\bigl(\frac{1}{E\tau}\bigr)},
\end{equation}
where we used Eqs.~\eqref{eq:zero_mode} and \eqref{eq:zero_mode_adj} to evaluate the matrix element \footnote{\label{footnote:matrix_elements} There is a technical nuance in the evaluation of the matrix element. To get a correct result, one needs to include a Heaviside step function $\Theta(\ln(1/E\tau) - |w|)$ in the definition of $f_0$ and $f_0^t$, and extend the integration over $w$ to the whole real line. Derivative $\partial / \partial w$ in the definition of $\hat{V}$ [see Eq.~\eqref{eq:V_operator}] should also be applied to the step function. The same holds to the calculation of all matrix elements presented below.}.
The expression for the diffusion coefficient in Eq.~\eqref{eq:lyap_distr_equation} reads 
\begin{widetext}
\begin{equation}\label{eq:diff_coeff}
    {\cal D} = \frac{1}{2} - \sum_{n = 1}^{\infty}\Bigl\{\frac{(f_0^t|\hat{V}|f_n)(f_n|\hat{V}|f_0)}{\Lambda_n} + \frac{(f_0^t|\hat{V}|\tilde{f}_n)(\tilde{f}_n|\hat{V}|f_0)}{\Lambda_n}  - \widetilde{\Lambda}_n\frac{(f_0^t|\hat{V}|f_n)(\tilde{f}_n|\hat{V}|f_0)}{\Lambda^2_n}\Bigr\}
\end{equation}
\end{widetext}
It is straightforward to see that
$(f_0^t|\hat{V}|f_n) = 0$ for all $n \geq 1$. Therefore, only the middle term in the brackets in Eq.~\eqref{eq:diff_coeff} contributes to ${\cal D}$. The matrix elements needed to evaluate it are given by
\begin{subequations}
\begin{align}\label{eq:sum_for_D}
    (f_0^t|\hat{V}|\tilde{f}_n) &=-\sqrt{\frac{8}{\ln\bigl(\frac{1}{E\tau}\bigr)}},\\
    (\tilde{f}^t_n|\hat{V}|f_0) &= \sqrt{\frac{2}{{\ln^3\bigl(\frac{1}{E\tau}\bigr)}}}\cdot\bigl[1 - 2 (-1)^n\bigr].
\end{align}
\end{subequations}
Using these expressions together with Eq.~\eqref{eq:f_n_Lambda_n} for $\Lambda_n$, we find
\begin{equation}
\sum_{n=1}^{\infty}\frac{(f_0^t|\hat{V}|\tilde{f}_n)(\tilde{f}_n|\hat{V}|f_0)}{\Lambda_n} = \frac{1}{3}.
\end{equation}
Therefore,
\begin{equation}\label{eq:D_final}
    {\cal D} = \frac{1}{6}.
\end{equation}
We note that the diffusion coefficient is independent of $E$; factors of $\ln(1/E\tau)$ in Eq.~\eqref{eq:sum_for_D} cancel between numerator and denominator. {The diffusion coefficient determines the variance of $\lambda$ via the Einstein relation, $\langle\hspace{-0.05cm}\langle \lambda^2 \rangle\hspace{-0.05cm}\rangle = 2{\cal D} x / (v\tau)$. Taking $x = L$, we obtain an expression presented in Eq.~\eqref{eq:distr_props}.}

Equation \eqref{eq:lyap_distr_equation} for $\Pi_0(\lambda|x)$ can be readily solved. It yields for $x = L$
\begin{align}
\Pi_0(\lambda) &\equiv \Pi_0(\lambda|L )\notag\\
&= \frac{\exp\Bigl[-\frac{v\tau}{4{\cal D}L}{\Bigl(\lambda - {L}/\bigr[{v\tau \ln\bigl(\frac{1}{E\tau}\bigr)\bigl]}\Bigr)^2}\Bigr]}{\sqrt{4\pi {\cal D} L /  v \tau}}.
\end{align}
This is the main result of this Appendix. It shows that the Lyapunov exponent is distributed normally, with the mean and the variance given by Eq.~\eqref{eq:distr_props}.

\section{General expression for $l(E)$\label{sec:general_l_E}}
In this Appendix, we derive a general expression for localization length $l(E)$ that is applicable for arbitrary relation between $\Delta_0, E$, and $1/\tau$. 

According to the recipe of Sec.~\ref{sec:disorder}, to obtain $l(E)$, we need to find the steady state distribution function of $\vartheta$ [cf.~Eq.~\eqref{eq:fokker-planck_st}], and then use it to perform the averaging in Eq.~\eqref{eq:loc_length}. Let us start with the former task. Instead of working with $\vartheta$ directly, it is convenient to use a logarithmic variable $w$. The form of the variable change is different for $\vartheta \in [-\pi, 0]$ and $\vartheta \in [0, \pi]$:
\begin{subequations}\label{eq:param_full}
\begin{align}
    w &= \ln \tan \frac{\vartheta}{2},\quad &\vartheta \in [0, \pi],\\
    w &= \ln \Bigl(-\cot \frac{\vartheta}{2}\Bigr),\quad &\vartheta \in [-\pi,0].
\end{align}
\end{subequations}
It maps each of the two intervals of $\vartheta$ onto a real line $w \in (-\infty, +\infty)$. 
We denote the part of the distribution function corresponding to $\vartheta \in [0, \pi]$ by $P_{+}(w)$, and the part corresponding to $\vartheta \in [-\pi, 0]$ by $P_-(w)$. It follows from Eq.~\eqref{eq:fokker-planck_st}, that functions $P_\pm$ satisfy the following Fokker-Planck equations:
\begin{equation}\label{eq:FP_w_two_trenches}
    v\tau \frac{\partial P_\pm}{\partial x} = 2\frac{\partial^2 P_\pm}{\partial w^2} - 2\tau\frac{\partial}{\partial w} \bigl([E \cosh w \pm \Delta_0] P_\pm\bigr).
\end{equation}
We are interested in the steady state solutions $P_{\pm}(w)$ of these equations. The steady state solutions are characterized by a constant probability current
\begin{equation}\label{eq:current_pm}
    {\cal J} = -\frac{2}{v\tau}\frac{\partial P_{\pm}}{\partial w} + \frac{2}{v}\frac{\partial}{\partial w}\bigl([E\cosh w\pm\Delta_0]P_\pm\bigr),
\end{equation}
which is the same for ${ P}_+(w)$ and ${ P}_-(w)$. 
Assuming $E > 0$ for concreteness (the localization length is even in $E$ due to the particle-hole symmetry) and solving Eq.~\eqref{eq:current_pm}, we find
\begin{equation}\label{eq:distribution_exact}
    P_{\pm}(w) = {\cal J} \frac{v\tau}{2} \int_0^{+\infty}\hspace{-0.2cm} e^{E\tau (\sinh w - \sinh (w + u)) \mp \Delta_0 \tau u} du.
\end{equation}
The value of the probability current ${\cal J}$ can be found from the normalization condition
\begin{equation}
    \int_{-\infty}^{+\infty}(P_+(w) + P_-(w))\,dw = 1. 
\end{equation}
We obtain with the help of the Nicholson's formula (see Eq.~(6.664.4) in Ref.~\onlinecite{gradstein_ryzhik}):
\begin{equation}\label{eq:prob_current}
    {\cal J} = \frac{2}{\pi^2 v \tau} \bigl[J_{|\Delta_0| \tau}^2(E\tau) + Y^2_{|\Delta_0|\tau}(E\tau)\bigr]^{-1},
\end{equation}
where $J_\nu(z)$ and $Y_\nu(z)$ are Bessel and Neumann functions of order $\nu$, respectively. Equations~\eqref{eq:distribution_exact} and \eqref{eq:prob_current} give an exact integral representation for the distribution function of the backscattering phase.

The found distribution function can be used to evaluate the localization length. Using Eq.~\eqref{eq:loc_length} together with Eq.~\eqref{eq:param_full}, we obtain an expression for $l(E)$ in terms of $P_\pm (w) \equiv P_{\pm} (w | E \tau, \Delta_0 \tau)$. The expression can be cast in the scaling form of Eq.~\eqref{eq:scaling_law}, with the scaling function
\begin{align}\label{eq:l_E_exact}
    &F(\varepsilon, \delta) =\notag\\
    &\Bigl[2\sum_{s = \pm} \int \Bigl( \frac{1}{\cosh^2 w} + s \delta \tanh w \Bigr) P_s(w| \varepsilon, \delta) dw\Bigr]^{-1}.
\end{align}
By computing integrals over $u$ and $w$ in Eqs.~\eqref{eq:distribution_exact} and \eqref{eq:l_E_exact} numerically, one can find $l(E)$ at arbitrary $E$, $\Delta_0$, and $1 / \tau$ (as long as the low-energy description is valid). This is how we obtained the curves in Fig.~\ref{fig:summary}(e) and a solid curve in Fig.~\ref{fig:disorder_numerics}(b).

\subsection{Energy-independence of $l(E)$ at $|\Delta_0| \tau = 1/2$\label{sec:special_point}}
Here we demonstrate that the localization length becomes independent of energy at $|\Delta_0| \tau = 1 / 2$. To do this, it is convenient to represent $l(E)$ as
\begin{align}
    &\frac{v\tau}{l(E)} = \frac{1}{2}\notag\\
    &+ \int dw du \, e^{E \tau (\sinh w - \sinh(w + u))}\bigl\{(1-2\tanh^2 w)  \notag\\
    &\times \cosh \frac{u}{2} - \tanh w \sinh \frac{u}{2}\bigr\},
\end{align}
where we used Eqs.~\eqref{eq:distribution_exact} and \eqref{eq:l_E_exact} with $|\Delta_0|\tau = 1 / 2$. In fact, the integral in the second and third lines is zero for any value of $E \tau$. One can verify this by performing an (asymptotic) series expansion of this integral in powers of $1 / E \tau$. It follows from a straightforward calculation that every term in the series vanishes. Thus, we find $l(E) = 2v\tau$, independently of $E$, see Fig.~\ref{fig:summary}(d).

\subsection{Density of states $\nu(E)$\label{sec:dos_derivation}}
The derived expression for the probability current [Eq.~\eqref{eq:current_pm}] allows one to find the density of states (DOS) of the wire $\nu(E)$. To see this, we note that the integrated DOS $N(E) = \int_0^{E} \nu(E^\prime) dE^\prime$ can be related to the backscattering phase winding number \cite{brouwer2011}:
\begin{equation}
    N(E) = \frac{1}{L} \frac{\vartheta(E,L)-\vartheta(0, L)}{2\pi} = \frac{1}{L} \frac{\vartheta(E,L)}{2\pi}.
\end{equation}
Here we used $\vartheta(0, L) = 0$, which follows from the particle-hole symmetry [cf.~Eqs.~\eqref{eq:ph_majoranas} and \eqref{eq:S_param}]. In its turn, the winding number equals (half) the number of times the $w$-variable passes through $w = +\infty$ in its evolution with $x$. On average, this number is
\begin{equation}
    \frac{\vartheta(E, L)}{2\pi} = {\cal J}L.
\end{equation}
Using the definition of $N(E)$ and Eq.~\eqref{eq:prob_current} for the probability current, we find \cite{ovchinnikov1977}
\begin{equation}\label{eq:dos_full}
    \nu(E) = \frac{2}{\pi^2 v \tau}\frac{d}{dE} \bigl[J_{|\Delta_0| \tau}^2(E\tau) + Y^2_{|\Delta_0|\tau}(E\tau)\bigr]^{-1}.
\end{equation}
Using the small-argument asymptotes of Bessel and Neumann function, $E \ll |\Delta_0|$, we obtain Eq.~\eqref{eq:dos_low_energy}. 
It also follows from Eq.~\eqref{eq:dos_full} that the dependence of DOS on $E$ becomes flat at the boundary of the critical region,  $|\Delta_0|\tau = 1/2$; DOS takes a disorder-independent value $\nu(E) = (\pi v)^{-1}$.

\section{Derivation of Eq.~\eqref{eq:loc_away_above} via application of Eqs.~\eqref{eq:fokker-planck_st} and \eqref{eq:loc_length}\label{sec:l_E_above_general}}

Here we apply general equations~\eqref{eq:fokker-planck_st} and \eqref{eq:loc_length} to obtain $l(E)$ at the above-the-gap energies, $E > \Delta_0$, for a wire detuned from the critical point by $\Delta_0 \gg 1/\tau$.
It follows from Eqs.~\eqref{eq:fokker-planck_st} and \eqref{eq:loc_length} that to find $l(E)$ we need to (i) obtain the distribution function of the backscattering phase $P(\vartheta)$, and (ii) compute the harmonics $\cos\vartheta$ and $\sin^2 \vartheta$ of the found $P(\vartheta)$.

Under the assumption that $E - \Delta_0 \gg E_{\star}$ [see Eq.~\eqref{eq:E_star} and related discussion for the definition of $E_{\star}$], the main term in Eq.~\eqref{eq:fokker-planck_st} is the drift term.
The phase $\vartheta$ rapidly slides down the washboard potential  $U_{\rm eff}(\vartheta) = -E\vartheta + \Delta_0 (\cos \vartheta - 1)$ in its evolution with $x$ [Eq.~\eqref{eq:backscattering}]; this motion is only weakly perturbed by the diffusion.
In the leading approximation, we can neglect the diffusion all together. Then we find the following normalized steady state distribution function:
\begin{equation}\label{eq:unpert_distr}
    P_{0}(\vartheta) = \frac{\sqrt{E^2 - \Delta_0^2}}{2\pi}\frac{1}{E + \Delta_0 \sin \vartheta}.
\end{equation}
We will also need a correction to $P_0(\vartheta)$ due to the diffusion. This correction is subleading in a small parameter $1/ \Delta_0 \tau \ll 1$, and can be obtained by treating the first term in Eq.~\eqref{eq:fokker-planck_st} as a perturbation. We find $P(\vartheta) = P_0(\vartheta) + P_1(\vartheta)$ with
\begin{equation}
    P_1(\vartheta) = \frac{\sqrt{E^2 - \Delta_0^2}}{2\pi}\frac{1}{\tau} \frac{E \sin \vartheta \cos \vartheta}{(E + \Delta_0 \sin \vartheta)^3}.
\end{equation}
We now use the obtained distribution function to evaluate the harmonics of $\vartheta$ in Eq.~\eqref{eq:loc_length}.
To the lowest non-vanishing order in $1 / \Delta_0 \tau$, we find for the two required harmonics:
\begin{subequations}\label{eq:two_harmonics}
    \begin{align}
    \langle \sin^2 \vartheta \rangle &= \frac{E}{\Delta_0}\Bigl(E/\Delta_0 - \sqrt{E^2/\Delta_0^2 - 1}\Bigr),\label{eq:sin_sq_0}\\
    \langle \cos \vartheta \rangle &= \frac{1}{\Delta_0 \tau} \cdot \frac{E}{\Delta_0}\Bigl(-\frac{1}{2} \frac{E \Delta_0}{E^2 - \Delta_0^2}\notag\\
    &\hspace{1.5cm}+ E/\Delta_0 - \sqrt{E^2/\Delta_0^2 - 1}\Bigr).\label{eq:cos_1}
    \end{align}
\end{subequations}
Although the latter harmonic is smaller than the former one by a factor $1 / \Delta_0 \tau \ll 1$, the contributions that the two produce to $1/l(E)$ are of the same order. Indeed, the coefficient of the first term in Eq.~\eqref{eq:loc_length} is larger than that of the second term by $\Delta_0 \tau \gg 1$, which compensates for the relative smallness of $\langle \cos \vartheta \rangle$. 
Using Eqs.~\eqref{eq:sin_sq_0} and \eqref{eq:cos_1} in Eq.~\eqref{eq:loc_length}, we recover Eq.~\eqref{eq:loc_away_above} for $l(E)$ obtained earlier with the help of Fermi Golden rule.

\section{Derivation of Eq.~\eqref{eq:l_subgap} via application of Eqs.~\eqref{eq:fokker-planck_st} and \eqref{eq:loc_length}\label{sec:l_E_subgap_general}}
In this Appendix, we use the general Eqs.~\eqref{eq:fokker-planck_st} and \eqref{eq:loc_length} to derive $l(E)$ at subgap energies.
At $E < \Delta_0$, the backscattering phase $\vartheta$ becomes trapped (in the course of its evolution with $x$, see Eq.~\eqref{eq:backscattering}) in a minimum of the washboard potential $U_{\rm eff}(\vartheta) = - E \vartheta + \Delta_0 (\cos \vartheta - 1)$. The phase leaves its trap only via the exponentially rare processes of the overbarrier ``thermal'' activation with ``temperature'' $\sim 1 / v\tau$. For $\Delta_0 - E \gg E_{\star}$, such processes can be neglected in the evaluation of $l(E)$.
The trapping happens at $\vartheta$ satisfying
\begin{equation}
    \sin \vartheta = E / \Delta_0.
\end{equation}
Using this expression in Eq.~\eqref{eq:loc_length}---and neglecting the second term $\sim 1 / v \tau \ll 1 / l(E)$ in the latter equation---we obtain $l(E) = v /\sqrt{\Delta_0^2 - E^2}$. This expression constitutes Eq.~\eqref{eq:l_subgap}.

\bibliography{references.bib}
\end{document}